\documentclass[journal]{IEEEtran}
\usepackage{textcomp}
\usepackage{listings}
\usepackage{multirow}
\usepackage{cite}
\ifCLASSINFOpdf
   \usepackage[pdftex]{graphicx}
   \DeclareGraphicsExtensions{.pdf,.jpeg,.png}
\else
\fi
\usepackage[cmex10]{amsmath}
\usepackage{amsthm}
\usepackage{amssymb}
\usepackage{algorithm}
\usepackage{algpseudocode}

\algnewcommand\algorithmicinput{\textbf{Parameter:}}
\algnewcommand\Param{\item[\algorithmicinput]}
\usepackage{array}
\usepackage{caption}
\usepackage{subcaption}

\usepackage{fixltx2e}
\usepackage{url}
\usepackage{balance}
\newtheorem{theorem}{Theorem}%[subsection]%[section]
\newtheorem{proposition}{Proposition}%[subsection]%[section]
\newtheorem{corollary}{Corollary}%[subsection]%[section]
\newtheorem{lemma}{Lemma}[theorem]
\newcommand*{\ROOT}{.}

\newenvironment{alphafootnotes}
{\par\edef\savedfootnotenumber{\number\value{footnote}}
	
	\setcounter{footnote}{0}}
{\par\setcounter{footnote}{\savedfootnotenumber}}

\begin{document}
\title{Guided-Processing Outperforms Duty-Cycling for Energy-Efficient Systems}


%\author{\IEEEauthorblockN{Long Le, Douglas L. Jones}
%\IEEEauthorblockA{Department of Electrical and Computer Engineering\\
%University of Illinois at Urbana-Champaign}}


\author{Long~N.~Le,~\IEEEmembership{Student Member,~IEEE,}
        and~Douglas~L.~Jones,~\IEEEmembership{Fellow,~IEEE}% <-this % stops a space
\thanks{L. Le and D.L. Jones are with the Department
of Electrical and Computer Engineering, University of Illinois at Urbana-Champaign, Illinois,
IL, 61801 USA.  D.L. Jones is currently Director of the Advanced Digital Sciences Center.}% <-this % stops a space
%\thanks{Manuscript received March 20, 2015; revised March 31, 2015.}
}


%\markboth{Journal of \LaTeX\ Class Files,~Vol.~13, No.~31, March~2015}%
%{Shell \MakeLowercase{\textit{et al.}}: Bare Demo of IEEEtran.cls for Journals}

\maketitle

\begin{abstract}
%\boldmath
Energy-efficiency is highly desirable for sensing systems in the Internet of Things (IoT). A common approach to achieve low-power systems is duty-cycling, where components in a system are turned off periodically to meet an energy budget. However, this work shows that such an approach is not necessarily optimal in energy-efficiency, and proposes \textit{guided-processing} as a fundamentally better alternative. The proposed approach offers 1) explicit modeling of performance uncertainties in system internals, 2) a realistic resource consumption model, and 3) a key insight into the superiority of guided-processing over duty-cycling. Generalization from the cascade structure to the more general graph-based one is also presented.
%can be viewed as a extension to the existing literature on cascade architectures.
%Further analyses shows that the cascade design itself is optimal, and a comprehensive comparison with an alternative approach, the duty-cycling design, shows that the cascade can uniformly outperforms the former. 
Once applied to optimize a large-scale audio sensing system with a practical detection application, empirical results show that the proposed approach significantly improves the detection performance (up to $1.7\times$ and $4\times$ reduction in false-alarm and miss rate, respectively) for the same energy consumption, when compared to the duty-cycling approach.
%is capable of continuous acoustic monitoring with an operational lifetime of one month on two D-cell batteries, which would otherwise be infeasible.
\end{abstract}


\begin{IEEEkeywords} 
	Guided-processing, duty-cycling, energy-efficient systems, resource-aware optimization, IoT.
\end{IEEEkeywords}

% \fi
\IEEEpeerreviewmaketitle

% Need to cite at least one reference for the compilation to be successfull.

\section{Introduction}\label{sec:intro}

Cisco predicted that by 2020, there will be 50 billions Internet-connected devices, ushering in the Internet of Things (IoT) paradigm \cite{atzori2010internet}. Many of these devices will be sensors that autonomously collect data about the physical world. Along with supporting infrastructures such as databases and data-analytics/inference engines, the resulting \textit{sensing system} is projected to enable many novel data-driven applications. While the new paradigm has much potential, it also comes with challenges. Among them are the `volume' and `velocity' \cite{mcafee2012big} 
of the data that need processing. It is becoming evident that the naive approach of stream-all-the-data-to-the-cloud is too costly in term of resources. And since energy is the most valuable resource in the post-Moore-law era, it is the target of interest for this work.

A straightforward approach to reduce the energy consumption of a sensing system is \textit{duty-cycling}, i.e.\ sensors are periodically turned off to reduce the amount of data that needs processing. While this approach does result in a low-power system, it does not necessarily yield an \textit{energy-efficiency} one, since the inference performance was completely ignored. An alternative approach is to have sensor nodes detect information-rich data instances from a data stream before uploading to the cloud for further processing. Unlike duty-cycling, this approach not only reduces the data-load, but also guides the downstream processing toward more quality data (hence the name \textit{guided-processing}). For instance, data streams from audio sensors contain mostly background noise, which can be screened out early by sensors. Intuitively, guided-processing solves the issue of duty-cycling by explicitly accounting for the inference performance, \textit{in addition} to the energy consumption.
%To address this problem, we believe it is important to decentralize the data processing on the cloud by offloading part of it to sensor nodes. By putting the processing close to the data source, 
%However, sensor nodes are highly resource-constrained and their processing algorithm must trade-off between its resource (e.g. energy) consumption and detection performance.

\begin{figure}
	\centering
	\includegraphics[width=\linewidth]{\ROOT/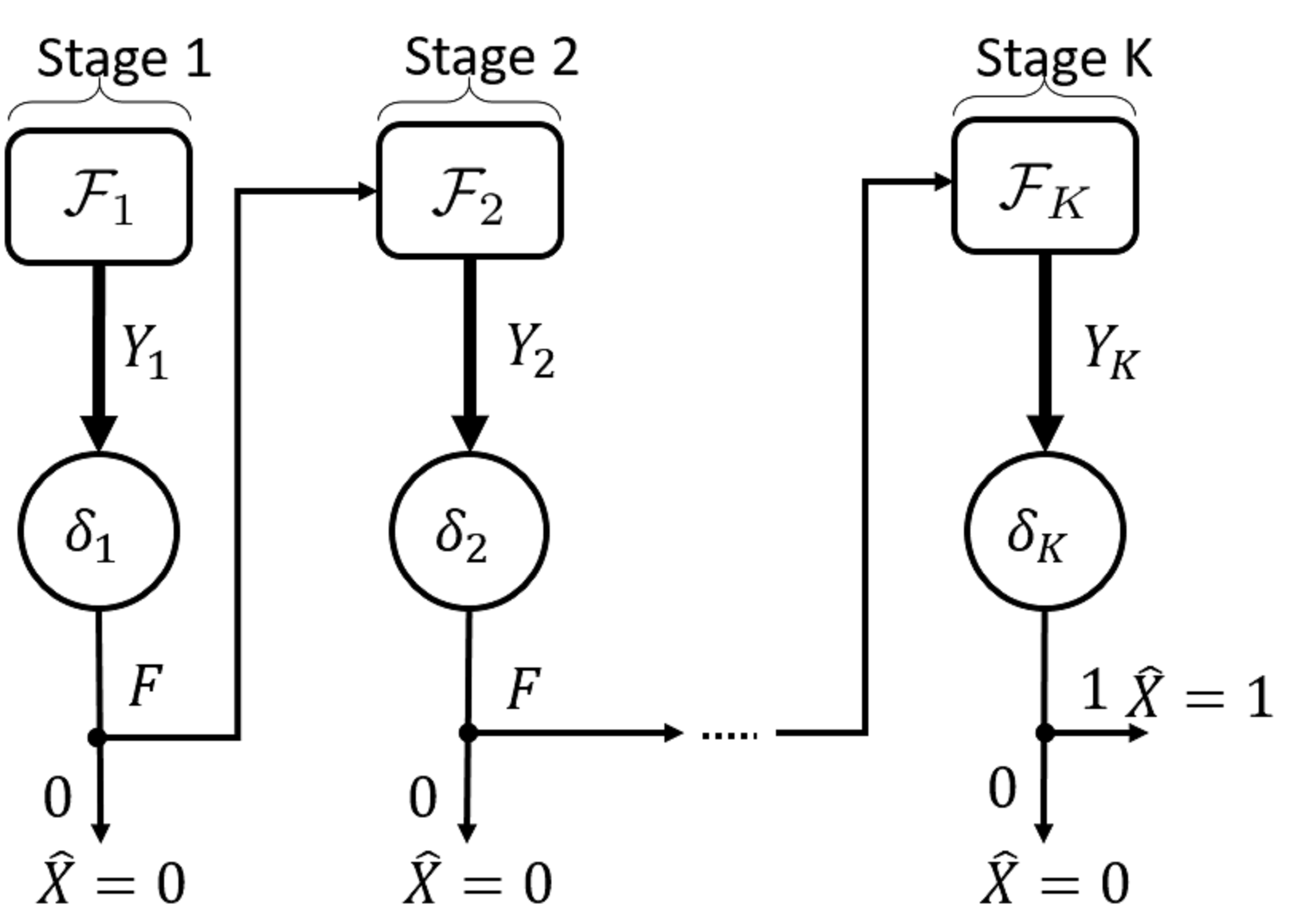}
	\caption{The cascade detection system with $K$ stages (indexed by subscripts). For stage $i$, $\mathcal{F}_i$ denotes the feature extractor and $\delta_i$ denotes the binary decision of a detector. The feature itself is denoted by $Y_i$. $X$ is the (detection) target's status, and $\hat{X}$ is the prediction about $X$ by a detector.}
	\label{fig:cascade}
\end{figure}

An architecture of a detection system that implements the guided-processing approach can be visualized in Fig.\ \ref{fig:cascade}. The system is a cascade of detection modules/detectors, each of which occupies a stage. A detector at stage $i$ consists of a feature extractor $\mathcal{F}_i$, which produces the feature $Y_i$, and a decision rule $\delta_i$, which takes $Y_i$ and all previous features $Y_1,\dots,Y_{i-1}$ as input. $\delta_i$ outputs different values depending on the stage (see Eq.\ \eqref{eqn:delta1}). $X$ is the (detection) target state, which takes value $1$ when the target is present, and $0$ otherwise. Finally, $\hat{X}$ denotes the prediction of $X$ by the detector.

The detection decision at each stage $\delta_i$ can take on the following values.
\begin{equation}\label{eqn:delta1}
\begin{aligned}
\delta_i &= 
\begin{cases}
0: \text{ stop and declare } \hat{X}=0 \text{ (negative)}\\
F: \text{ extract feature next}\\
\end{cases}\\
&i = 1,\dots,K-1\\
\delta_K &= 
\begin{cases}
0: \text{ declare } \hat{X} = 0 \text{ (negative)}\\
1: \text{ declare }\hat{X} = 1 \text{ (positive)}
\end{cases}
\end{aligned}
\end{equation}
Note that only negative decisions, i.e.\ $\hat{X} = 0$, are allowed at intermediate stages ($i =1,\dots,K-1$) since the goal is not to make the final decision (which is reserved for the last stage with the best performance) but to screen out early negative instances that are more likely in a rare-target setting.

The cascade architecture has been studied before in the literature. For instance, the seminal work by Viola and Jones \cite{viola2001rapid} showed empirically that such a design is very effective in detecting rare targets in a large dataset (e.g.\ face detection), and was also proposed as a resource-efficient approach for stream mining by Turaga et al.\ in \cite{turaga2006resource}. Detailed comparisons with existing works are articulated in Section \ref{sec:related}. Our contributions here include the explicit modeling of performance uncertainties at intermediate layers/stages of a system. In addition, a realistic resource/energy consumption model is proposed. Finally, an in-depth comparison with the duty-cycling approach reveals a key insight on how a guided-processing system uniformly outperforms a duty-cycling one in term of energy-efficiency. Furthermore, it is worth noting that the proposed principle is more general than both the tandem/cascade and the parallel (for instance, see \cite{le2013energy}) structures, and can be applied to more sophisticated ones like trees and graphs to create inference-aware, low-power sensing systems.
%For instance, 
%\begin{itemize}
%	\item Our resource-consumption model is more accurate than existing works, which often suffers from a `nebulous' resource-consumption model that is inapplicable in practice.
%	\item Our performance model also accounts for uncertainty in feature models at intermediate stages of the cascade, an approach that has not been explored before.
%	\item Beside optimizing parameters of the cascade, we also show that, under mild conditions, the cascade design itself is optimal. That is, adding additional degrees of freedom such as early positive decisions to the cascade structure does not improve its performance.
%	\item Another popular, low-power design is duty-cycling, in which the system cycles between the on and off states \cite{dutta2005design}. Our formulation can provide a detailed analysis of the performance different between the proposed cascade and the duty-cycling approaches.
%\end{itemize}
%It is worth noting that the proposed cascade abstraction is applicable for both centralized and distributed implementation. Namely, the modules of the cascade can be software components running on the same hardware, or distributed nodes across the network.

The rest of the paper is organized as follows. Section \ref{sec:related} reviews prior works that studied the cascade structure, along with the limitations of their formulations/solutions. Section \ref{subsec:featModel} sets up preliminaries for the system model presented in Section \ref{subsec:sysOpt}, where the method to optimize its operation is also discussed. The analytical comparison between the optimal cascade system and the duty-cycling one is given in Section \ref{subsec:compareDuty}. Guided-processing on graph is presented in Section \ref{sec:graph_gp}. In Section \ref{sec:sim}, the proposed theory was applied to the design of an energy-efficient acoustic sensing system. Final remarks are given in Section \ref{sec:concl}.

%%%%%%%%%%%%%%%%%%%%%%%%%%%%%%%%%%%%%%%%%%%%%%%%%%%%
%%%%%%%%%%%%%%%%%%%%%%%%%%%%%%%%%%%%%%%%%%%%%%%%%%%%

\section{Related works}\label{sec:related}

It is worthwhile to note that the cascade detection system of interest here is different from the serial detector network in the distributed detection literature \cite{tang1991optimization,swaszek1993performance,viswanathan1988optimal}, in which the decision of a current module is treated as an extra observation, instead of as a control signal to \textit{censor} subsequent modules and conserve resources.

The cascade architecture is prevalent in many inference applications, with the most widely-known example being the seminal work in face detection by Viola and Jones \cite{viola2001rapid}. In \cite{viola2001rapid}, the system of cascaded detection modules is used to quickly discard many negative sub-images typically observed in face-detection applications, thus significantly speeding up the detection process. However, the cascade is not optimized in \cite{viola2001rapid}, leaving the optimal classifiers' parameters, both thresholds and weights, to be desired.

To this end, Luo \cite{luo2005optimization} proposed to optimize thresholds of each detection module in a cascade using the classical Neyman-Pearson criterion, without consideration of resource cost. Under the assumption of statistical independence between detection modules, a gradient-based algorithm is proposed to search for the locally optimal thresholds, which is also a limitation of \cite{luo2005optimization}. In contrast, our approach guarantees a globally optimal, resource-aware solution and does not assume independence between stages.

Later, Jun and Jones \cite{jun2010energy} incorporated an energy resource constraint in the Neyman-Pearson-based optimization over thresholds of a two-stage cascade. In this setting, three solution types were identified: one that utilizes all of the available energy and false-alarm rate, one that utilizes all the energy while slacking the false-alarm constraint, and one that utilizes all the false alarm while slacking the energy constraint. An algorithm to find the optimal thresholds is only available if the true solution is of the first type. Later, it is proven in \cite{jun2013cascading} that, if observations of the first stage are reused/resampled in the second stage, then the first-type solution is optimal. Furthermore, the individual performance of the first and second stage detectors were used as the lower and upper bounds on the (detection) performance of the cascade, respectively. However, there was no comparison with the duty-cycling approach in term of energy-efficiency. Finally, unlike \cite{jun2010energy,jun2013cascading}, whose goal is the design of energy-efficient sensor nodes (for which a two-stage architecture is often sufficient), this work undertakes the design of an entire sensing system (for which there are likely more than one downstream processing). The new setting therefore motivates the development of a more general solution, i.e.\ cascade systems with an arbitrary number of stages.

Cho et al.\ \cite{chen2016dual} proposed a two-stage cascade architecture for an ultra low-power acoustic sensor. The first stage coarsely samples time-frequency (TF) characteristics of an audio stream and triggers the full TF analysis in the second stage if an acoustic event is detected. However, there was no attempt at optimizing the triggering threshold.

Chen et al.\ \cite{chen2012fidelity} designed a surveillance system using a two-stage cascade of low-end (acoustic and infrared) and high-quality (camera) sensors. The system in \cite{chen2012fidelity} can find a triggering threshold that either minimizes the detection error, or satisfies a constraint on the CPU utilization for video processing, but not both, and a heuristic was used to combine the two solutions, i.e.\ use the threshold that minimizes the detection error if it also satisfies the utilization constraint, otherwise use the one that satisfies the constraint. Unlike the ad-hoc approach of \cite{chen2012fidelity}, our solution is derived from a well-defined framework.
It is worth noting that Cohen et al.\ \cite{cohen2013managing} also studied a similar problem in which a multi-modal sensing system (with a PIR sensor and a camera) was designed for monitoring vehicles. While the treatment in \cite{cohen2013managing} is principled (based on the partially-observable Markov decision process (POMDP) framework), the sensors are \textit{not} operated in cascade, but instead are equally plausible options at each time step, and hence is different from our work.

Since the optimization of the cascade is hard,
Raykar et al.\ \cite{raykar2010designing} relaxed the problem by 
assuming classifiers in the cascade produce soft/probabilistic 
outputs instead of hard decisions, and converted the joint optimization of
classifiers' linear weights into a maximum {\it a posteriori} problem.
Feature costs are also incorporated into the optimization using the
standard Lagrangian argument, and a gradient-based algorithm is used to
find the optimal weights. However, the thresholds must be found using
an exhaustive grid search, which is computationally intensive
for cascades with many classifiers. Our solution does not suffer this drawback.

Chen et al.\ \cite{chen2012classifier} proposed a cyclic optimization algorithm to optimize the linear weights of the classifiers in the cascade, along with their early-exit thresholds. That is, at each iteration, the algorithm cycles through all classifiers in the cascade, optimizing each one while leaving others untouched. The algorithm stops when the loss function no longer improves. A disadvantage of such optimization procedure is that it requires multiple passes through the cascade, and there is no theoretical bound on the number of iterations it will take. In contrast, our solution requires only a single pass through the cascade.

In stream mining, Turaga et al.\ \cite{turaga2006resource} employed a cascade of Gaussian mixture model (GMM)-based classifiers and formulated a problem to find both the number of mixture components and the threshold in each classifier that maximize the system detection rate subject to constraints on false alarm, memory and CPU. The solution in \cite{turaga2006resource} takes a person-by-person approach where it iterates between 1) finding optimal numbers of mixture components, i.e.\ resource allocation, for all classifiers given thresholds, and 2) finding optimal thresholds for a given resource allocation. However, this approach failed to capture the direct dependence of the cascade's resource consumption on its thresholds, and is inherently suboptimal.
%and later extension by Fu et al.\ \cite{fu2007configuring}

A limitation of the above works is that they only considered open-loop solutions where the thresholds are independent variables to be optimized. Ertin \cite{emre1999polarimetric} considered closed-loop solutions for the two-stage cascade detection problem where the optimal decision rule at each stage, which is observation-dependent, is sought. It was shown that the optimal policies are still likelihood ratio tests, but with coupling thresholds, i.e.\ the threshold at a stage depends on the receiver operating characteristic (ROC) and the threshold of the other stage. Namely, the optimal thresholds can not be found using the solution technique employed by \cite{emre1999polarimetric}. Note that, unlike classical detection problems, optimizing thresholds in a cascade is critical in the trade-off between inference performance and resource cost. A contribution of this paper is finding the optimal parameters (both test-statistics and thresholds) for general detection systems.

Trapeznikov et al.\ studied a generalization of the cascade that was termed multi-stage sequential reject classifier (MSRC), which is simply the cascade with an additional positive decision \cite{trapeznikov2013multi} or multiple additional (classification) decisions\cite{trapeznikov2013supervised} at intermediate stages. Their resource-consumption model is `nebulous', i.e.\ if the decision at an intermediate stage is to defer to the next stage, an abstract, \textit{independent} "penalty" is incurred. In contrast, in our resource model, these penalties are shown to be precisely the Lagrangian-weighted of the feature extraction costs, and hence they are coupled (see Eq.\ \eqref{eqn:VMultiApp}).

On the other hand, a resource-consumption model closely related to ours was considered by Wang et al.\ in \cite{wang2014lp}. The minor difference is that, instead of being proposed, our model was derived from first principles. However, \cite{wang2014lp} formulated the problem using the empirical risk minimization framework, since it was assumed that probabilistic models of high-dimensional features cannot be estimated. We take a different approach where it is assumed that probabilistic models of features \textit{can} be estimated, by first reducing the features' dimensionality. In other words, the inputs into our algorithm are (probabilistic) models, not a dataset as in \cite{wang2014lp}. In addition, the solution proposed in \cite{wang2014lp} is a convex linear-program, which requires a convex relaxation (with an upper-bounding convex surrogate function) of the true objective function. In contrast, our solution is a dynamic program and requires no relaxation.

%%%%%%%%%%%%%%%%%%%%%%%%%%%%%%%%%%%%%%%%%%%%%%%%%%%%
%%%%%%%%%%%%%%%%%%%%%%%%%%%%%%%%%%%%%%%%%%%%%%%%%%%%
\section{Optimality analysis of a cascade detection system}\label{sec:cascade}

\subsection{Feature models}\label{subsec:featModel}
For the rest of the document, the colon notation is used to denote a collection, e.g.\
\begin{equation}
y_{1:i} \triangleq \{ y_1, \dots, y_{i-1}, y_i\}
\end{equation}

%\subsubsection{Probabilistic models}
Recall that $Y_i$ denotes the feature used by the detector at stage $i$, and is modeled as a random variable whose distribution depends on the latent target $X \in \{0,1\}$, i.e.\
\begin{equation}\label{eqn:probModel}
\begin{aligned}
%Y_1 &\sim \mathrm{p}_1(y_1 | x),  x \in \{0,1\}\\
%Y_i &\sim \mathrm{p}_i(y_i | x,y_{1:i-1}),  x \in \{0,1\}, i = 1,\dots, K
Y_i &\sim \mathrm{p}_i(y_i | x),  x \in \{0,1\}, i = 1,\dots, K
\end{aligned}
\end{equation}
where lower-case letters denote realizations of the corresponding random variable in upper case and $\mathrm{p}$ denotes a probability mass/density function. It is assumed that these distributions are \textit{stationary} and hence can be estimated during training. While the stationary assumption might seems too restrictive at first glance, it does not preclude practical implementations of subsequent results, as will be shown in Proposition \ref{prop:nonstatic}.
Finally, it is worth noting that the feature (conditional) distributions in \eqref{eqn:probModel} are chosen by Nature and thus \textit{conditionally independent} of prior stages' decisions (if any), given the target state.
The decisions do influence the \textit{belief} about the latent state though.

Using Bayes' rule, the posterior probability of target presence is given by
\begin{equation}\label{eqn:genModel}
\begin{aligned}
&\pi_1(y_1) = \frac{1}
{ 1+\frac{1-\pi_0}{l_1(y_1)\pi_0} }\\
&\pi_i(y_{1:i}) = \frac{1}
{ 1+ \frac{1- \pi_{i-1}(y_{1:i-1})}{l_i(y_i) \pi_{i-1}(y_{1:i-1})} }\\
&\hspace{2cm}i = 2,\dots, K
\end{aligned}
\end{equation}
where $l_i(y_i) \triangleq \mathrm{p}_i(y_i|1)/\mathrm{p}_i(y_i|0)$ and $\pi_i(y_{1:i}) \triangleq \mathrm{P}(X=1|y_{1:i})$ are the likelihood function and posterior probability at stage $i$, respectively. $\pi_0 \triangleq \mathrm{P}(X=1)$ is the prior probability of the target presence. Finally, the evidence probability is given by
\begin{equation}\label{eqn:evidenceGen}
\mathrm{p}_{i}(y_{i}|y_{1:i-1}) = \mathrm{p}_{i}(y_{i}|1)\pi_{i-1} + \mathrm{p}_{i}(y_{i}|0)(1-\pi_{i-1})
\end{equation}

An important aspect of the cascade detection system is that, except for the last stage, the main goal of other stages is to quickly screen out negative instances, and not to make the final decision. Therefore features used at stages other than the last one are suboptimal for the detection task by  design, to keep the cost of their execution low. For instance, the all-band energy feature can neither characterize a bandpass target precisely, nor distinguish between a bandpass target and another bandpass interference, but can still be useful in the cascade thanks to its low cost \cite{jun2013cheap}. The sub-optimality of these early-stage features, either due to 1) the failure to discriminate the target against potential interferences, or 2) the insufficient modeling of the target, can all be modeled as \textit{uncertainty} in feature models. To this end, we employ the following \textit{least-favorable} feature density models, developed by Huber in the context of robust detection \cite{huber1968robust},\cite[Chapter 10]{huber2011robust},\cite[Chapter 6]{levy2008principles}, in place of the nominal ones.
\begin{equation}\label{eqn:leastFavor}
\begin{aligned}
\mathrm{p}_i(y|0) &\leftarrow 
\begin{cases}
\frac{1-\epsilon_{0i}}{v'+w'l_{Li}} [v'\mathrm{p}_i(y|0) + w'\mathrm{p}_i(y|1)], l_i(y)<l_{Li}\\
(1-\epsilon_{0i}) \mathrm{p}_i(y|0), l_{Li}\leq l_i(y)\leq l_{Ui}\\
\frac{1-\epsilon_{0i}}{w''+v''l_{Ui}} [w''\mathrm{p}_i(y|0) + v''\mathrm{p}_i(y|1)], l_i(y)>l_{Ui}
\end{cases} \\
\mathrm{p}_i(y|1) &\leftarrow
\begin{cases}
\frac{(1-\epsilon_{1i})l_{Li}}{v'+w'l_{Li}} [v'\mathrm{p}_i(y|0) + w'\mathrm{p}_i(y|1)], l_i(y)<l_{Li}\\
(1-\epsilon_{1i}) \mathrm{p}_i(y|1), l_{Li}\leq l_i(y)\leq l_{Ui}\\
\frac{(1-\epsilon_{1i})l_{Ui}}{w''+v''l_{Ui}} [w''\mathrm{p}_i(y|0) + v''\mathrm{p}_i(y|1)], l_i(y)>l_{Ui}
\end{cases}\\
&i = 1,\dots,K-1
\end{aligned}
\end{equation}
where the `$\leftarrow$' symbol is the assignment operator and
\begin{equation}
\begin{aligned}
v' = \frac{\epsilon_{1i}+\nu_{1i}}{1-\epsilon_{1i}}, 
v'' = \frac{\epsilon_{0i}+\nu_{0i}}{1-\epsilon_{0i}}\\
w' = \frac{\nu_{0i}}{1-\epsilon_{0i}}, 
w'' = \frac{\nu_{1i}}{1-\epsilon_{1i}}
\end{aligned}
\end{equation}
and $0 \leq\epsilon_{0i},\epsilon_{1i},\nu_{0i},\nu_{1i} \leq 1$ are uncertainty parameters of stage $i$. $l_{Li}$ and  $l_{Ui}$ are the lower and upper bounds of the likelihood ratio at stage $i$, respectively, and can be found by solving the equations outlined in \cite[Chapter 6]{levy2008principles}. Note that since the new least-favorable densities result in a bounded likelihood function, the corresponding posterior probability is also bounded.
\begin{equation}\label{eqn:piRobust}
\begin{aligned}
\pi_{Li} \triangleq \frac{1}{1+\frac{1-\pi_{i-1}}{l_{Li}\pi_{i-1}}} \leq \pi_i(y_{1:i}) \leq 
\pi_{Ui} \triangleq \frac{1}{1+\frac{1-\pi_{i-1}}{l_{Ui}\pi_{i-1}}}
\end{aligned}
\end{equation}

%\subsubsection{Computational models}
%Alternatively, computational models can be used instead of probabilistic models. In particular, logistic regression is used to model the conditional probability, i.e.\
%\begin{equation}\label{eqn:discModel}
%\begin{aligned}
%\pi_i(y_{1:i}) = \frac{1}{1+\exp(-\sum_{j=1}^i\beta_j y_j - \alpha)}
%\end{aligned}
%\end{equation}
%where $\beta_j, j = 1,\dots, i$ and $\alpha$ are parameters of the regression that can be learned from a training data set $\{(y(n),x(n)), n = 1,\dots,N\}$  of size $N$ for each detector. Clearly, \eqref{eqn:discModel} is equivalent to \eqref{eqn:genModel} when
%\begin{equation}
%\begin{aligned}
%\alpha &= \ln\frac{\pi_{i-1}}{1-\pi_{i-1}}\\
%\beta_i &= \frac{\ln l_i(y_i|y_{1:i-1}) - \sum_{j=1}^{i-1}\beta_j y_j}{y_i},\\
%&i = 1,\dots,K
%\end{aligned}
%\end{equation}
%The evidence probability is given by
%\begin{equation}\label{eqn:evidenceDisc}
%\mathrm{p}_i(y_i|y_{1:i-1}) = \frac{\sum_{n=1}^N\mathbb{I}(y_i=y(n)|y_{1:i-1})}{N}
%\end{equation}
%where $\mathbb{I}()$ denotes the indicator function that takes value $1$ when its argument is true and $0$ otherwise.

%Similar to the probabilistic case, computational feature models of early stages can account for uncertainty by ensuring that they are bounded between some $\pi_{Li}, \pi_{Ui},i=1,\dots,K-1$, i.e.\
%\begin{equation}\label{eqn:discRobust}
%\pi_i \leftarrow \pi_{Li} + \frac{\pi_{Ui}-\pi_{Li}}{1+\exp(-\sum_{j=1}^i\beta_j y_j - \alpha)},
%i = 1,\dots,K-1
%\end{equation}

%The results in the following sections are applicable for both probabilistic and computational models.

\subsection{System model and optimization}\label{subsec:sysOpt}
Optimizing the cascade system amounts to finding optimal decision rules $\delta_{1:K}$ that jointly minimize the proposed system's Bayes risk $R_B$ of incorrect decisions subject to an expected system resource (e.g.\ energy) constraint $e$.
\begin{equation}\label{eqn:mainObj}
\begin{aligned}
&\min_{\delta_{1:K}} R_B(\delta_{1:K})\\
&s.t.\  E(\delta_{1:K}) \leq e
\end{aligned}
\end{equation}
where $E$ is the expected system resource consumption. The Lagrangian technique can be used to convert the constrained optimization problem \eqref{eqn:mainObj} into the following unconstrained, yet regularized, one
\begin{equation}\label{eqn:minR}
\min_{\delta_{1:K}} R(\delta_{1:K}) \triangleq \lambda E + R_{K,\text{A}} + \sum_{i=1}^{K}R_{i,\text{M}}
\end{equation}
where the parameter $\lambda$, which depends on the resource constraint $e$, couples the resource consumptions of all stages together and $R$ denotes the \textit{system risk}, which is a measure of the combined detection performance and resource consumption. Hence, it is evident that a system with lower system risk is more energy-efficient. The Bayes risk $R_B$ has been broken down into multiple terms. $R_{i,M},i=1,\dots,K-1$ are the miss (false negative) risks due to early negative decisions at intermediate stages. $R_{K,M},R_{K,A}$ are the miss and false-alarm (false positive) risks due to incorrect decisions at the last stage. Note that the system has no false-alarm risk at intermediate stages, since the cascade structure does not allow early positive decisions to be made. There are two reasons for this. First, to a dummy detector that flips a coin to make decisions, rare target makes it more likely to incur a false-alarm than a miss. Second, intermediate stages with model uncertainties are also likely to be fooled by interference to trigger a false-alarm. Altogether it is relatively safe to ignore early positive decision, since they are too unreliable. Proposition \ref{prop:vivaCascade} later shows precisely when this ignorant is unharmful.

The expected resource consumption at stage $i$ is the resource cost of feature extraction, denoted by $D_i$, weighted by the probability of that feature being selected by the previous stage. In addition, even when features are \textit{not} extracted, real systems also incur a small, but non-zero, stand-by power consumption which is modeled by $d_i < D_i,i=2,\dots,K$. Hence,
\begin{equation}\label{eqn:lambdaE}
\begin{aligned}
E &\triangleq D_1 + \sum_{i=1}^{K-1} \left[ D_{i+1}\mathrm{P}(\delta_i = F) +d_{i+1}\mathrm{P}(\delta_i = 0) \right]\\ 
\end{aligned}
\end{equation}
where $D_1$ is weighted by $1$ because the first-stage feature is always extracted. Lastly, $D_i$ and $d_i$ can be measured in practice by profiling the feature-extraction process, and resource costs generally go up by an order of magnitude\footnote{It is noteworthy that this exponential cost increase is similar to that considered by Poor in the context of quickest change detection \cite{poor1998quickest}, where it is shown that the optimal statistic is still the well-known accumulated likelihood product, but additionally weighted by the exponential base at each iteration.} from one layer to another.

The solution to Problem \eqref{eqn:minR} is given by the following theorem.

\begin{theorem}\label{thm:optDecRules}
	(The optimal decision rules for the cascade)
	\begin{equation}\label{eqn:optDecRules}
	\begin{aligned}
	\delta_i^{\ast}(\pi_i) &= 
	\begin{cases}
	0, \pi_i(y_{1:i}) < \tau_i^{\ast}\\
	F, \text{ else}
	\end{cases}\\
	&i = 1,\dots,K-1\\
	\delta_K^{\ast}(\pi_K) &= 
	\begin{cases}
	0, \pi_K(y_{1:K})  < \tau_K^{\ast}\\
	1,  \text{ else}
	\end{cases}\\
	\end{aligned}
	\end{equation}
	
	where $\tau_i^{\ast} \in [\pi_{Li},\pi_{Ui}]$ are the optimal thresholds at stage $i$.
	
	\begin{proof}
		See Appendix \ref{subsec:app1}.
	\end{proof}
\end{theorem}

Eq.\ \eqref{eqn:optDecRules} in Theorem \ref{thm:optDecRules} shows that the posterior probabilities of intermediate stages can be used to guide the execution of subsequent stages by thresholding them to decide whether to stop or extract more features in the next stage.
The final stage has a standard detection rule, with the posterior probability being thresholded to make a prediction about the target state. The optimal threshold values \{$\tau_i^{\ast}$\}, which are critical in this trade-off between performance and resource cost, can be found using Algorithm \ref{alg:recur_thresh}.

\begin{algorithm}[t]
	\caption{Pseudo-code to find optimal thresholds for the cascade system.}\label{alg:recur_thresh}
	\begin{algorithmic}[1]
		\Function{optimize}{$\textit{model}$}
		\State $\textit{model}$ is a structure containing the system's feature models
		\State $M$ is the probability quantization size
		\State $b=[0:1/(M-1):1]$ (dummy) probability vector 
		\State Use \eqref{eqn:leastFavor}
		%(for probabilistic models) or fit \eqref{eqn:discRobust} with training data (for computational models)
		to obtain robust versions of $\textit{model}$.
		\State $V_K = \min(C_{M}b, C_{A}(1-b))$
		\State $\tau_K^{\ast}$ = $C_{A}/(C_{A}+C_{M})$
		\For {$i = K-1:-1:1$}
		\State $J$ = expected next-stage ($i$+1) value function
		\State $V_i = \min(C_{M}b+\lambda{\bf d}_{i+1}, J)$
		\State $\tau_i^{\ast} = \min \{b: V_{i} - (C_{M}b+\lambda{\bf d}_{i+1}) < 0\}$ 
		\State $\tau_i^{\ast} = \max(\pi_{Li},\min(\pi_{Ui},\tau_i^{\ast}))$
		\EndFor
		\State $J$ = expected next-stage (1) value function
		\State $V_0 = J$
		\EndFunction
		%
		%\algstore{myalg}
	\end{algorithmic}
\end{algorithm}

The solution offered by Theorem \ref{thm:optDecRules} requires the conversion of a feature $Y$ into a posterior $\pi$, which can be difficult for practical implementations. To address this issue, an alternative, adaptive form of the solution, similar to the one proposed in \cite[Eq.\ (14)]{jun2013cascading}, is given as follows.
\begin{proposition}\label{prop:nonstatic}
	(Adaptive implementation)
	Let
	\begin{equation}\label{eqn:adaptive}
	\begin{aligned}
	\delta_i^{\ast}(y_i) &= 
	\begin{cases}
	0, y_{i} < \eta_i\\
	F, \text{ else}
	\end{cases}\\
	&i = 1,\dots,K-1\\
	\delta_K^{\ast}(y_K) &= 
	\begin{cases}
	0, y_{K}  < \eta_K\\
	1,  \text{ else}
	\end{cases}\\
	\end{aligned}
	\end{equation}
	where the adaptive thresholds $\eta_i$ are updated as follows.
	\begin{equation}
	\eta_i \leftarrow \eta_i + \mu(\hat{q}_i - q_i), i = 1,\dots,K
	\end{equation}
	with
	\begin{equation}
	\begin{aligned}
	q_i &\triangleq \mathrm{P}(\pi_i \geq \tau_i^\ast|\pi_{i-1}), i = 1,\dots,K-1\\
	q_K &\triangleq \mathrm{P}(\pi_K \geq \tau_K^\ast|\pi_{K-1})
	\end{aligned}
	\end{equation}
	being the \textit{activation probabilities} and $\hat{q}_i, i=1,\dots,K$ are their runtime estimates. Finally, $\mu$ is the adaptation step size. Then \eqref{eqn:adaptive} is equivalent to \eqref{eqn:optDecRules}, provided that the features' likelihood ratios are monotonic.
\end{proposition}
The advantage of the adaptive form in Proposition \ref{prop:nonstatic} is that it does not require runtime posterior evaluations, but instead $K$ probability functions (of the prior $\pi_{i-1}$ of stage $i$) $q_i,i=1,\dots,K$ that can be computed at train time. Intuitively, the thresholds $\eta_i$ in this implementation are updated to ensure that \eqref{eqn:adaptive} produces decisions with the same probability measure $q_i$ as that of \eqref{eqn:optDecRules}, consequently making them equivalent (assuming all features have monotonic likelihood ratios).

Given the above optimal decisions, the Corollary \ref{corol:perf} quantifies the corresponding performance of the cascade system.

\begin{corollary}\label{corol:perf}
	(Optimal performance of the cascade)
	\begin{equation}
	R^{\ast}(\pi_0) \triangleq R(\delta_{1:K}^{\ast},\pi_0) = V_0(\pi_0)
	\end{equation}
	where $V_0(\pi_0)$ is the result of the following recursion
	\begin{equation}\label{eqn:Vrecur}
	\begin{aligned}
	V_K(\pi_K) &\triangleq \min (\underbrace{C_M\pi_K}_{\text{miss risk}}, \underbrace{C_{A}(1-\pi_K)}_{\text{false-alarm risk}}),\pi_K\in[0,1]\\
	V_{i}(\pi_{i}) &\triangleq \min (C_M\pi_i + \lambda{\bf d}_{i+1},\\
	&\hspace{3em} \underbrace{\lambda D_{i+1} + \mathbb{E}[V_{i+1}(\pi_{i+1}(Y_{i+1},\pi_{i}))]}_{\text{expected next-stage value function}} ),\\
	&\pi_i\in [\pi_{Li},\pi_{Ui}], i = 1,\dots, K-1\\
	V_{0}(\pi_{0}) &\triangleq \lambda D_{1} + \mathbb{E}[V_{1}(\pi_{1}(Y_{1},\pi_{0}))] \\
	\end{aligned}
	\end{equation}
	And the corresponding optimal thresholds are given by
	\begin{equation}\label{eqn:thresh}
	\begin{aligned}
	\tau_K^{\ast} &= C_{A}/(C_{A}+C_{M})\\
	\tau_i^{\ast} &= \max(\pi_{Li},\min(\pi_{Ui},\\
	&\hspace{1em} \min \{\pi_i: V_{i}(\pi_{i}) - \left[C_M\pi_i + \lambda {\bf d}_{i+1}\right] < 0\})),\\
	&i = 1,\dots, K-1\\
	\end{aligned}
	\end{equation}
	where $C_{M},C_{A}$ are the costs associated with miss and false-alarm decisions and
	\begin{equation}
		{\bf d}_{i} \triangleq \sum_{j=i}^K d_{j}
	\end{equation} 
	is the (backward) accumulated off-costs.
\end{corollary}

Corollary \ref{corol:perf} shows that the optimal performance achieved by the system can be found using a recursive procedure. The procedure has $K$ iterations, each corresponding to a stage in the system. Starting from the last stage $K$ and proceeding backward to $0$, the value function $V$ is recursively updated (see \eqref{eqn:Vrecur}). The last-stage value function $V_K$ is the minimum of the miss and false-alarm risk across $\pi_K$. An intermediate-stage value function $V_i,i=1,\dots,K-1$ is the minimum of the miss risk and the \textit{expected next-stage} value function, which requires the probabilistic updates in Eq.\ \eqref{eqn:genModel},\eqref{eqn:evidenceGen}.
%(for probabilistic models) or \eqref{eqn:discModel},\eqref{eqn:evidenceDisc} (for computational models). 
The final value function $V_0$ is the minimal risk achievable by the system.

Once a value function is known, then the corresponding optimal threshold can be found using just arithmetic operations, i.e.\ comparing the value function with the combined miss risk and off-mode resource consumptions ${\bf d}_{i}$. For the last stage $K$, the optimal threshold can be given in closed form. Note that the intermediate stages' thresholds are capped between the upper and lower bounds due to model uncertainty (see Section \ref{subsec:featModel}).

The discussion so far has been focusing on optimizing parameters of the cascade design. A natural next question is whether the constraints of the cascade design can be relaxed to further improve performance. Namely, would introducing additional degrees of freedom, i.e.\ early positive decisions in intermediate stages, to the cascade \textit{always} improve its performance? Intuitively, when model uncertainties of intermediate stages are accounted for (see Section \ref{subsec:featModel}), and it is known \textit{a priori} that the target is rare, early positive decisions are likely to have higher risk and hence are discouraged. Therefore, introducing additional early positive decisions does \textit{not} always improve the performance of the cascade. The precise conditions for which the cascade design itself is optimal is given by the following proposition.
\begin{proposition}\label{prop:vivaCascade}
	(Optimality of the cascade design)
	With model uncertainty, introducing additional early positive decisions in intermediate stages of the cascade does not improve performance, when
	\begin{equation}\label{eqn:condVivaCascade}
	\begin{aligned}
	\underbrace{\max\{\pi_i: V_i(\pi_i) - \left[ C_{A}(1-\pi_i)+\lambda{\bf d}_{i+1} \right] < 0   \}}_{\text{optimal threshold for early positive decision}} > \pi_{Ui},\\
	i = 1,\dots,K-1
	\end{aligned}
	\end{equation}
	
	\begin{proof}
		See Appendix \ref{subsec:app2}.
	\end{proof}
\end{proposition}

The left-hand side of \eqref{eqn:condVivaCascade} is the optimal threshold corresponding to an early positive decision. Namely, these additional decisions also have threshold-based optimal policies (see Appendix \ref{subsec:app2}), and a posterior probability \textit{above} such a threshold shall trigger an early positive decision. If such a threshold is above the upper bound on the posterior probability at a stage, then its early positive decision is never selected, and hence does not affect the performance of the cascade.

\subsection{Guided-processing vs duty-cycling}\label{subsec:compareDuty}

As alluded to in Section \ref{sec:intro}, duty-cycling is an alternative low-power design in which the system switches between the \textit{on} and \textit{off} modes. The duration for the on mode is determined by the duty-cycle factor $\rho \in [0,1]$, with $\rho = 1$ being always on and $\rho=0$  being always off. When off, the system completely misses out any potential events. However, when on, the system uses the best feature model, i.e.\ an equivalent of the cascade's last stage. Hence, the duty-cycling design can be viewed as the extreme version of the cascade without intermediate layers.
%(see Fig.\ \ref{fig:dutyCycling}). 
The (Bayes) detection risk and the resource consumption of a duty-cycled system is therefore given by
\begin{equation}\label{eqn:duty}
\begin{aligned}
R_{\text{dc},B} &= \rho(\underbrace{R_{\text{dc},M} + R_{\text{dc},A}}_{\text{risk in the on mode }}) + (1-\rho)\underbrace{C_M\pi_0}_{\substack{\text{miss risk in} \\ \text{the off mode}}}\\
E_{\text{dc}} &= \rho D_{\text{dc}} + (1-\rho)d_{\text{dc}}
\end{aligned}
\end{equation}
where $R_{\text{dc},M}$,$R_{\text{dc},A}$ are the miss and false-alarm risks during the on mode, respectively. $D_{\text{dc}}$, $d_{\text{dc}}$ are the resource consumptions in the on and off modes, respectively. 

In general, $D_{\text{dc}} \geq D_K$ and $d_{\text{dc}} \geq d_K$ because they include not only the resource consumption of the last stage, but also additional overhead needed to get the data there. In addition, $R_{\text{dc},B} \geq R_{K,B} \triangleq R_{K,M}+R_{K,A}$ (see Appendix \ref{subsec:app4}). Hence, the theoretically best duty-cycling system is the one in which the above bounds are met with equality.
 
Optimizing the duty-cycling design is straightforward since the detection risk and the resource consumption are decoupled. Hence the optimal detection rule does not affect the resource consumption, and $\rho$ can be adjusted to meet a resource budget. While the duty-cycling design has the advantage of being simple, it can result in a lower energy-efficiency compared to the cascade design. Indeed, Proposition \ref{propo:1} shows that the optimal cascade design can outperform even the best duty-cycling design uniformly (across all $\rho\in[0,1]$, for a given $\pi_0$).

\begin{proposition}\label{propo:1}
	(Guided-processing vs duty-cycling) The optimal cascade design outperforms the best duty-cycling design uniformly (across all duty-cycle factor $\rho\in [0,1]$), provided that
	\begin{equation}\label{eqn:condBetter1}
	R^\ast(\pi_0) \leq C_{M}\pi_0+\lambda d_{K}
	\end{equation}
	and
	\begin{equation}\label{eqn:condBetter2}
	\underbrace{\sum_{i=1}^{K-1}R_{i,M}^\ast(\pi_0)}_{\text{intermediate-stages' miss risk}} \leq \underbrace{\lambda(D_K - e)}_{\text{weighted resource saving}}
	\end{equation}
	where
	\begin{equation}\label{eqn:RiM}
	\sum_{i=1}^{K-1}R_{i,M}^\ast(\pi_0) \triangleq V_{0,M}(\pi_0)
	\end{equation}
	and $V_{0,M}(\pi_0)$ is the result of the following recursion
	\begin{equation}\label{eqn:ViM}
	\begin{aligned}
	V_{K,M}(\pi_K) &= 0,\pi_K\in[0,1]\\
	V_{i,M}(\pi_{i}) &= \begin{cases}
	C_M\pi_i, &\pi_i\leq \tau_i^\ast\\
	\mathbb{E}[V_{i+1,M}(Y_{i+1},\pi_{i})], &\text{ else }\\
	\end{cases},\\
	&\pi_i\in [\pi_{Li},\pi_{Ui}], i = 1,\dots, K-1\\
	V_{0,M}(\pi_{0}) &= \mathbb{E}[V_{1,M}(Y_{1},\pi_{0})] \\
	\end{aligned}
	\end{equation}
	
	\begin{proof}
		See Appendix \ref{subsec:app4}.
	\end{proof}
\end{proposition}
Eq.\ \eqref{eqn:condBetter1} is simply a sanity check to ensure that the minimal risk of the proposed design must be lower than that of doing nothing. Eq.\ \eqref{eqn:condBetter2} is more involved and it highlights the core differences between the proposed and duty-cycling approaches. In term of detection performance, the guided-processing approach fundamentally incurs more miss risk (i.e.\ additional miss terms) due to the introduction of intermediate stages, i.e.\ the left-hand side of Eq.\ \eqref{eqn:condBetter2} and defined in Eq.\ \eqref{eqn:RiM}, to reduce the energy consumption. Hence, the key insight is, as long as the achieved resource saving, i.e.\ the right-hand side of Eq.\ \eqref{eqn:condBetter2}, is more than the additional miss risk incurred (for a given $\pi_0$), then the guided-processing design uniformly outperforms even the theoretically best duty-cycling one.

\section{Graph-based guided-processing}\label{sec:graph_gp}

Recall from the discussion on cascade structures (Section \ref{subsec:sysOpt}) that the guided-processing solution starts from the last detector and works backward to the first one, since each stage depends on the value function of a downstream stage. Therefore, to adapt the established solution to graph-based systems, a \textit{post-order traversal} through nodes is required, since each node, which herein represents a detection module, depends on value functions of its downstream neighbors. An obvious technical requirement is that there must be no cycle in the system's (directed) graph, i.e.\ only directed-acyclic graphs (DAG) are admissible. For instance, a post-order traversal on the graph in Fig.\ \ref{fig:vGraph,chap:graph} is $9 \rightarrow 6 \rightarrow 7 \rightarrow 8 \rightarrow 2 \rightarrow 3 \rightarrow 4 \rightarrow 5 \rightarrow 1$. Note that while there are more than one valid post-order traversals, they are all equivalent from the guided-processing's perspective.

\begin{figure}
	\centering
	\includegraphics[width=\linewidth]{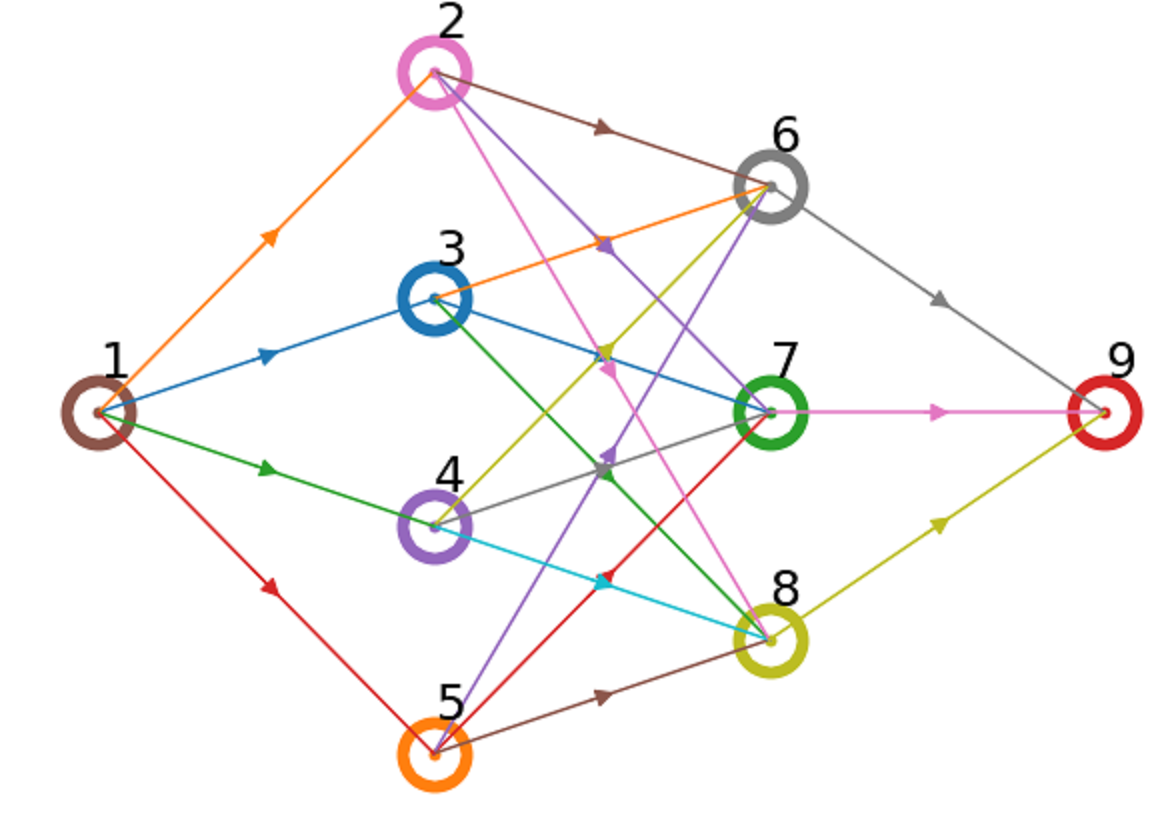}
	\caption{A sample graph-based detection system where each node is a module. There are $K=9$ modules in the system, labeled accordingly.}
	\label{fig:vGraph,chap:graph}
\end{figure}

At each node/iteration, with all downstream neighbors processed as the result of the post-order traversal, the guided-processing equations are given as follows.
\begin{equation}\label{eqn:recur,chap:graph}
\begin{aligned}
V_i(\pi_i) &= \min (C_M\pi_i,C_A(1-\pi_i)), \text{ if } \mathcal{N}(i) = \emptyset\\
V_i(\pi_i) &= \min (C_M\pi_i+\lambda {\bf d}_n, \{ \lambda D_n + \mathbb{E}[V_n(\pi_n(Y_n,\pi_i))]:\\
&n\in \mathcal{N}(i) \}), \text{ else }\\
V_0(\pi_0) &= \lambda D_1 + \mathbb{E}[V_1(\pi_1(Y_1,\pi_0))]
\end{aligned}
\end{equation}
where it is assumed that the miss and false-alarm cost of all nodes are the same and, again, denoted by $C_M, C_A$. Similarly, $D_n$ and ${\bf d}_n$ are the resource on-cost (for extracting feature $Y_n$) and the accumulated (from all downstream nodes) off-cost of node $n$. The symbols $i,V_i,\pi_i$ denote the current node, its value function, and its posterior probability, respectively. $\mathcal{N}(i)$ denotes the set of all (downstream) neighbors of node $i$. Nodes with no neighbor are last/terminal nodes.

The corresponding optimal decision functions are given as follows.
\begin{equation}\label{eqn:graphOptDec}
\begin{aligned}
\delta_i^\ast(\pi_i) &= 
\begin{cases}
0, \text{ if } V_i(\pi_i) = C_M\pi_i\\
1, \text{ else }
\end{cases}, \text{ if } \mathcal{N}(i) = \emptyset\\
\delta_i^\ast(\pi_i) &= 
\begin{cases}
\begin{aligned}
F_n, &\text{ if } V_i(\pi_i) = \lambda D_n + \\
&\mathbb{E}[V_n(\pi_n(Y_n,\pi_i))], n \in \mathcal{N}(i)
\end{aligned}\\
0, \text{ else }
\end{cases}, \text{ else }\\
\end{aligned}
\end{equation}
where $F_n$ is the decision to choose a (downstream) neighbor $n$ for subsequent feature extraction. For practical implementation, the solution in \eqref{eqn:graphOptDec} is rewritten in term of the generalized activation probability $q_i$ as follows.
\begin{equation}
\begin{aligned}
q_i(\pi_{i-1}) &\triangleq \{\mathrm{P}(\delta_i^\ast = a| \pi_{i-1}): a = 0,1\}, \text{ if } \mathcal{N}(i) = \emptyset\\
q_i(\pi_{i-1}) &\triangleq \{\mathrm{P}(\delta_i^\ast = a| \pi_{i-1}): a = 0,F_n, n\in \mathcal{N}(i)\}, \text{ else }\\
\end{aligned}
\end{equation}
where the notation $\pi_{i-1}$ is slightly abused to denoted the \textit{union} of all admissible priors of node $i$'s parents, i.e.\
\begin{equation}
\begin{aligned}
\pi_{i-1} &\triangleq \cup_{p \in \mathcal{P}(i)} \pi_p\\
&= [\min_{p\in \mathcal{P}(i)}\pi_{Lp}, \max_{p\in \mathcal{P}(i)}\pi_{Up}]
\end{aligned}
\end{equation}
where $\mathcal{P}(i)$ denotes the set of node $i$'s parents and $\pi_{Lp}, \pi_{Up}$ are the lower and upper bounds on the admissible prior at a parent node $p$.
The union operation follows from the assumption that triggers from parent nodes are mutually exclusive.

\section{System prototype}\label{sec:sim}

This section applies the theory developed in Section \ref{sec:cascade} to design an energy-efficient audio sensing system.
%The expected operational lifetime is one month on 2 D-cell batteries, each has 1700 mAh at 3.6V. Therefore, the average power (current) budget is $1700\times2\div 30\div 24 = 4.72$ mA.

\subsection{Hardware components}\label{subsec:hardware}

\begin{figure}[t!]
	\centering
	\includegraphics[width=\linewidth]{\ROOT/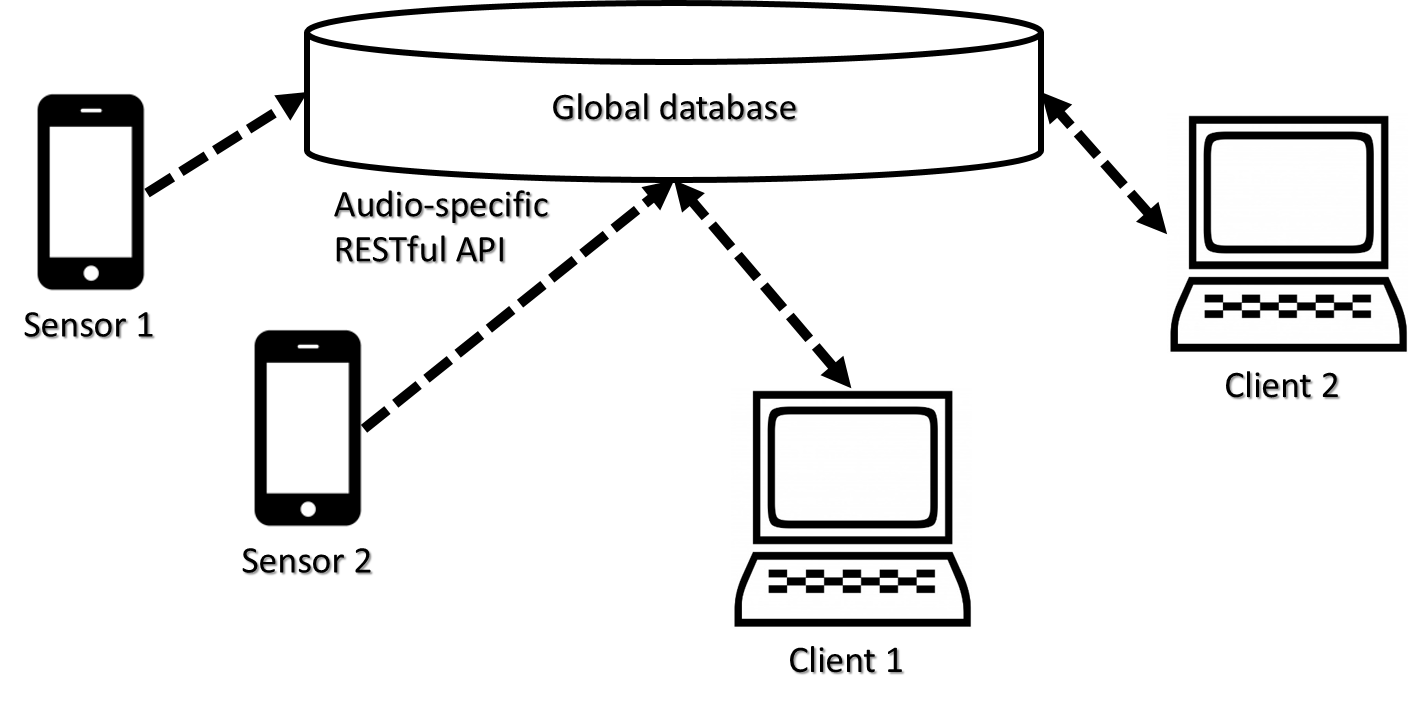}
	\caption{Devices of the prototype audio sensing system.}
	\label{fig:hardware}
\end{figure}

\begin{figure}
	\centering
	\fbox{\includegraphics[width=0.69\linewidth]{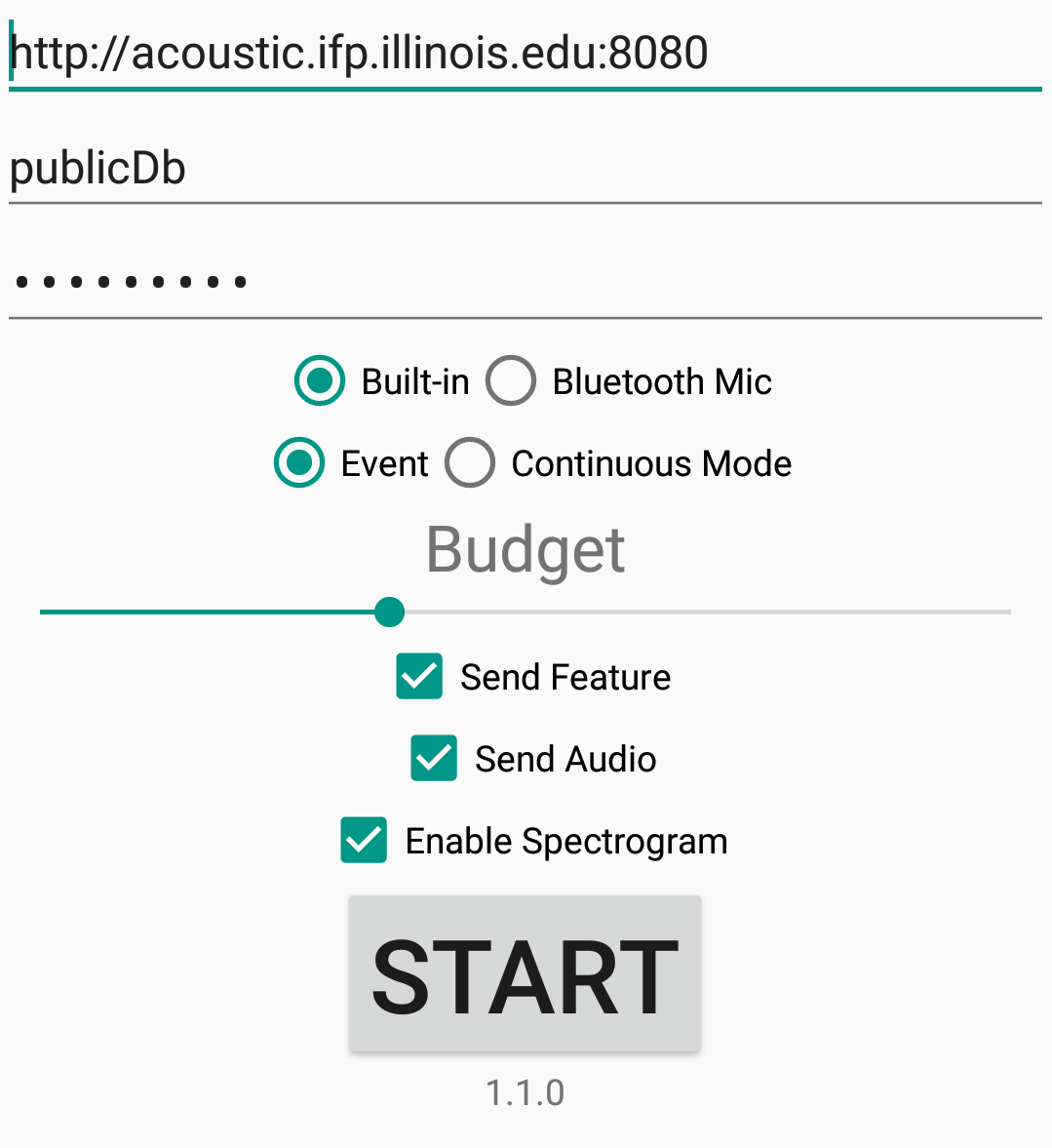}}
	\caption{A screenshot of the proposed Android-based audio analysis app. The app uses the adaptive implementation outlined by Proposition \ref{prop:nonstatic}, with the probability $q_1$ input via the ``Budget'' slider.}
	\label{fig:android}
\end{figure}

The proposed sensing system consists of three classes of devices: sensors, clients, and a globally-accessible data-plane \cite{mor2016toward} (see Fig. \ref{fig:hardware}). In our current prototype, the data-plane is an instance of MongoDb database \cite{chodorow2013mongodb} with a custom RESTful interface specialized for audio data. Sensors are Android smartphones with our audio analysis app (See Fig.\ \ref{fig:android}) installed\footnote{Available for download at \url{https://play.google.com/store/apps/details?id=com.longle1.spectrogram}}. Finally, clients are standard PCs running the Windows OS. The power consumption of sensors, profiled using Trepn\cite{trepn} on a Nexus-5X, and clients, measured using \textit{powercfg} on a $2.00$ GHz machine, at different operating modes are listed in in Table \ref{tab:components}.
%The power consumption of the microphone, the preamp circuit, and the ADC, altogether is 1 mA and considered as the baseline of the system. That leaves a power budget of 3.72 mA for the processor. From Table \ref{tab:components}, it is clear that the above power constraint cannot be met with the processor always active, and the software component needs to manage the processor's power budget efficiently by exploiting its off mode.

\begin{alphafootnotes}
	\begin{table}[H]
		\caption {Power consumption at different modes of devices of the acoustic sensing system.} \label{tab:components} 
		\centering
		\begin{tabular}{|c|c|}
			\hline
			\textbf{Devices \& Modes} & \textbf{Power consumption (mW)} \\ 
			\hline
			%Android processing & 84.36\footnotemark[1] \\ 
			Android processing & 84.36\\
			\hline
			%Android transmission  & 1097\footnotemark[1]\\
			Android transmission  & 1097\\
			\hline
			%PC sleep & 264\footnotemark[2]\\
			PC sleep & 264\\
			\hline
			%PC processing & 15131\footnotemark[2]\\
			PC processing & 15131\\
			\hline
		\end{tabular}
	\end{table}
	%\footnotetext[1]{Profiled using Trepn\cite{trepn} on Nexus-5X.}
	%\footnotetext[2]{Profiled using powercfg under Windows 10 on a 2.00 GHz machine.}
\end{alphafootnotes}

%\begin{figure}[t!]
%	\centering
%	\begin{subfigure}[t]{0.5\textwidth}
%		\centering
%		\includegraphics[height=1.5in]{\ROOT/phdthesis/cascadePower}
%		\caption{The energy consumption of the cascade design for one period.}
%		\label{fig:cascadePower}
%	\end{subfigure}
%	~ 
%	\begin{subfigure}[t]{0.5\textwidth}
%		\centering
%		\includegraphics[height=1.5in]{\ROOT/phdthesis/dutyCyclingPower}
%		\caption{The energy consumption of the duty-cycling design for one period.}
%		\label{fig:dutyCyclingPower}
%	\end{subfigure}
%	\caption{}
%\end{figure}

\subsection{Software components}

While the proposed sensing system can be used for many applications, the detection of the Golden-cheeked Warbler (GCW)'s (type-A) calls \cite{leonard2010variation} is chosen here as the application of interest. Namely, $X = 1$ indicates the presence of a GCW call, and $X = 0$ otherwise. Since the GCW is an endangered bird species, this application has important implications for their conservation.

\begin{figure}
	\centering
	\includegraphics[width=\linewidth]{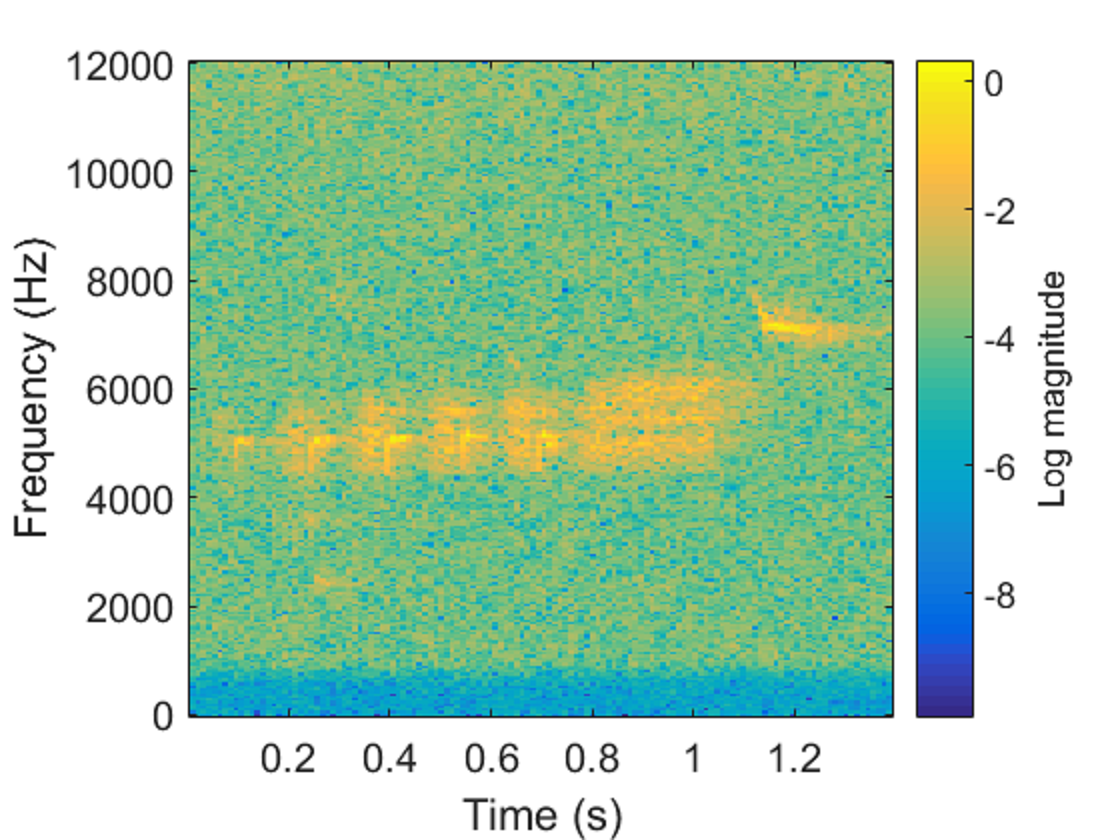}
	\caption{Spectrogram of a sample GCW's (type-A) call.}
	\label{fig:specgram}
\end{figure}

The application's software is organized into three subtasks: generic energy-based analysis, spectral-based analysis, and temporal-spectral-based analysis. The energy analysis is a low-complexity computation that produces energy-based features useful for detecting acoustic events from silence. The spectral-based analysis takes into account the spectral information about the GCW calls, which only has energy in the 4500-6500 Hz and 7000-8000 Hz bands (see Fig.\ \ref{fig:specgram}), to produce band-specific, energy-based features using standard DSP filtering techniques. Finally, the spectral-temporal-based analysis takes into account both the spectral and temporal structure of the GCW call from Fig.\ \ref{fig:specgram} to produce reliable, indicative features using a template matching technique. Note that the input into the above analyses is an audio stream (or precisely, its high-dimensional time-frequency representation, see Fig.\ \ref{fig:specgram}), and their output is a scalar score sequence, i.e.\ a score for each audio frame. Hence, these analyses effectively perform dimensionality reduction.

\begin{figure}[t!]
	\centering
	\includegraphics[width=\linewidth]{\ROOT/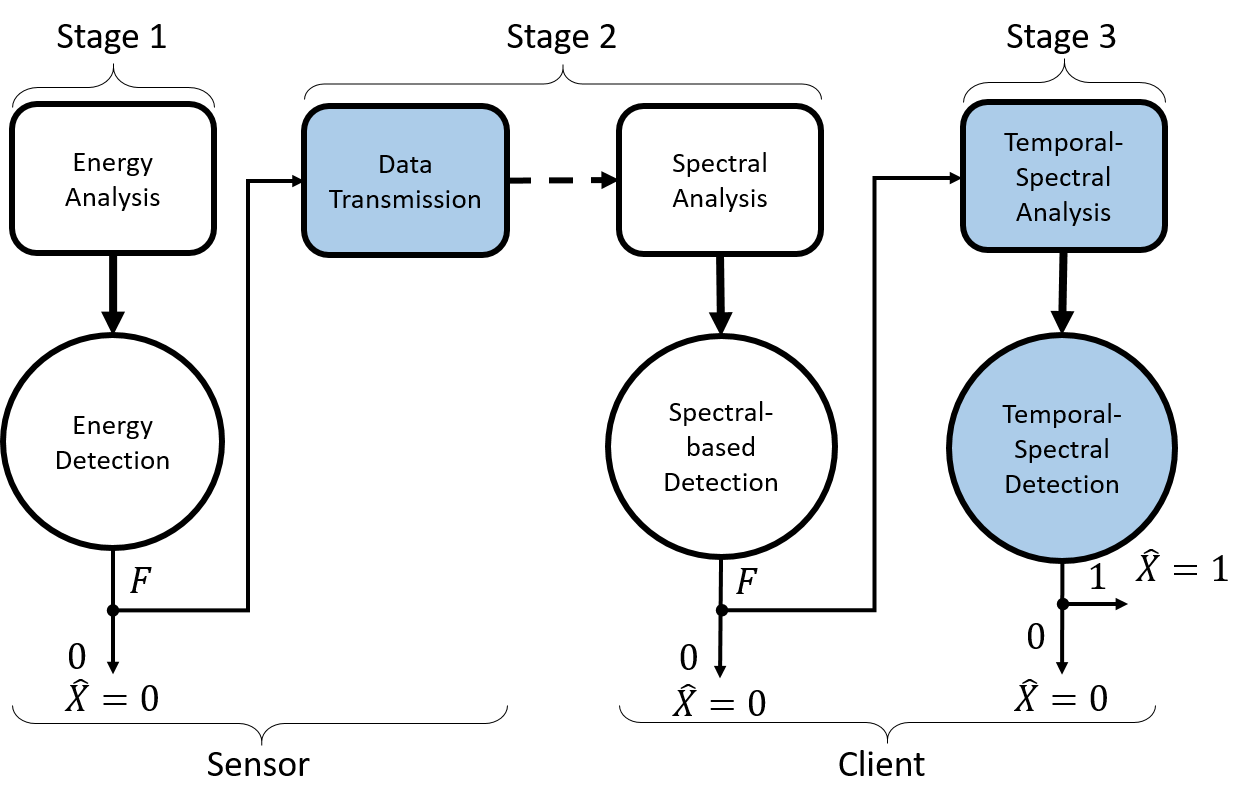}
	\caption{The software block diagram is organized as a cascade with 3 stages: energy analysis as stage 1, spectral-based analysis (along with the data transmission) as stage 2, and temporal-spectral analysis as stage 3. Note that components of the cascade are implemented distributedly across the network, with the dashed line representing a remote connection. For comparison, a system with the duty-cycling design only has highlighted components, i.e.\ data transmission from sensor to a client where the temporal-spectral analysis is carried out.}
	\label{fig:cascadeReal}
\end{figure}

Since the generic energy analysis has low computational complexity and can help prune out a significant amount of noise-only data from the audio stream early, it is executed on edge/sensor nodes. Only \textit{acoustic events} are transmitted downstream to clients, where spectral and temporal-spectral-based analyses are further carried out. The system diagram is illustrated in Figure \ref{fig:cascadeReal} and arranged to fit the proposed cascade abstraction. Note that the physical separation (between sensors and clients) does not necessarily correspond to the logical separation (between stages). For instance, the cost of data transmission on sensors are included into the cost of executing the second stage, along with the cost of spectral-based analysis on clients, since they are both a result of the first-stage decision.

The resource cost parameters at each stage $D_i, i=1,2,3$, which can be estimated from values of Table \ref{tab:components} and the execution times of the software components, are needed to optimize the resource-performance trade-off. It is assumed that all processing finishes before a periodic deadline, i.e.\ when buffers (an ADC buffer on the sensor, a task buffer on the client) are full. The average execution time of each task (per audio frame of $32$ ms) can be estimated/profiled and is given as follows. The energy analysis takes $16$ ms\footnote{Estimated as half of the frame length.}. The average transmission time takes $11$ ms ($500$ ms for a $1.5$ s event\footnote{Profiled on the Android prototype.}).
%(see Fig.\ \ref{fig:sensorPower}). 
Finally, the spectral and temporal-spectral analyses take $0.34$ $\mu$s and $14$ ms, respectively\footnote{Profiled in MatLab on the PC.}.
% (see Fig.\ \ref{fig:clientPower}). 
Hence,
\begin{equation}
\begin{aligned}
D_1 &= 84.36 \times 0.016,\\
D_2 &= 1097\times 0.011 + 15131\times 0.34\times 10^{-6}, \\
D_3 &= 15131\times 0.014,\\
\end{aligned}
\end{equation}
The off-mode/idle energy costs (per audio frame) on the client are given as follows.
\begin{equation}
\begin{aligned}
d_2 &= 264\times 0.34\times 10^{-6}\\
d_3 &= 264\times 0.014
\end{aligned}
\end{equation}

The same system designed with the duty-cycling approach will has less components (only those highlighted in Fig.\ \ref{fig:cascadeReal}) and its resource/energy consumptions parameters are given as follows.
\begin{equation}\label{eqn:D_dc}
\begin{aligned}
D_{\text{dc}} &= 1097\times 0.011 + 15131\times 0.014\\
d_{\text{dc}} &= 264\times 0.014
\end{aligned}
\end{equation}

Our dataset is a 46-minute, 24 kHz audio recording at the field in Rancho Diana, San Antonio's city park.
%The audio was originally recorded in 24 kHz, but is resampled to 16 kHz for all processing.
The dataset contains 206 GCW calls (manually identified and labeled), each of whose duration is approximately one second. In addition to GCW calls, the dataset also contains various interferences from other animals' vocalization, time-varying wind noise, etc., since it is taken directly from field recording. Precisely, the fraction of GCW calls in the entire dataset is 10.19\%. Hence, this detection problem belongs to the rare-target class, where the prior is asymmetrical, i.e.\ $\pi_0 \ll 0.5$. Throughout this section, we consider a range of prior in the rare-event regime, i.e.\ $\pi_0 \in [0.05,0.15]$. Finally, the miss and false-alarm costs are given by $C_M = 3, C_A = 1$ to emphasize that the miss risk is higher in this setting.

\begin{figure}
	\centering
	\includegraphics[width=\linewidth]{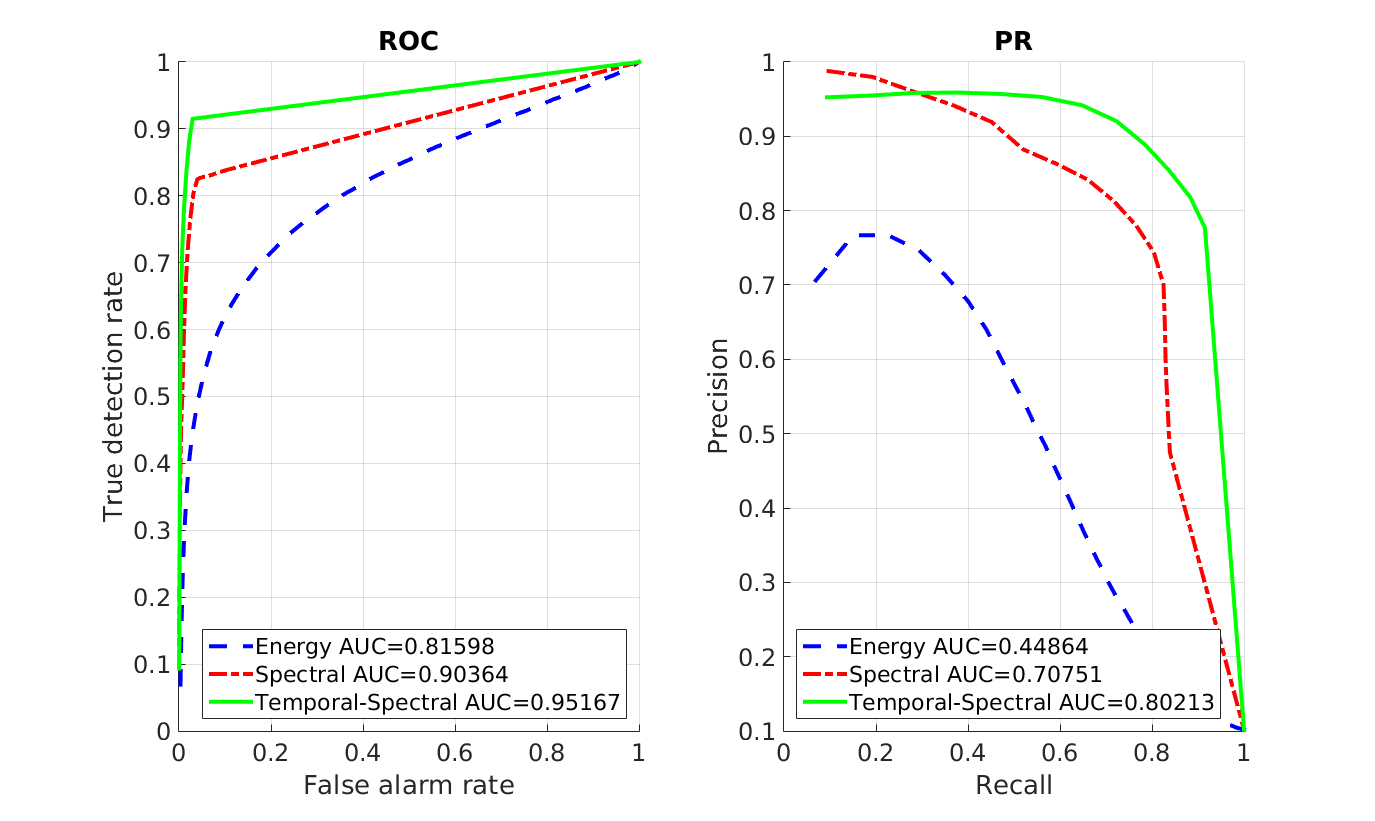}
	\caption{Receiver operating characteristic (ROC) curves and precision-recall (PR) curves of the features produced by the 3 analyses.}
	\label{fig:roc}
\end{figure}

The dataset are input to each of the three analyses discussed above. The scalar output scores from each analysis are taken as its respective features, resulting in a total of three feature sets/groups/types. The discriminative power of each feature type, or equivalently the performance of an analysis, can be quantified using receiver operating characteristic (ROC) and precision-recall (PR) curves as shown in Fig.\ \ref{fig:roc}. From the figure, it is evident that the temporal-spectral feature is better than the spectral feature, which in turn is better than the generic energy feature, at detecting GCW calls.

The conditional probability mass functions (PMF), i.e.\ $\mathrm{p}_i(y_i|x)$, of features from each analysis can be estimated up to some quantization level, i.e.\ $100$. 
%It is assumed that the PMF of each feature group is conditionally independent given the GCW labels. 
Furthermore, as alluded to in Section \ref{subsec:featModel}, energy-based and spectral-based features, by construction, are inadequate to characterize GCW calls, and hence there are inherent uncertainties in these features for the detection of GCW calls. These uncertainties can be explicitly accounted for in the features' distributions using the uncertainty model discussed in Section \ref{subsec:featModel}, with the following parameters.
\begin{equation}
\begin{aligned}
\epsilon_{01} &= \epsilon_{02} = 0.1\\
\epsilon_{11} &= \epsilon_{12} = 0.1\\
\nu_{01} &= \nu_{02} = 0.1\\
\nu_{11} &= \nu_{12} = 0.1\\
\end{aligned}
\end{equation}
Intuitively, the $\epsilon$ and the $\nu$ parameters indicate the level and the strength of a contamination on the nominal distribution, respectively. A formal method to set these parameters are left for future work.
Finally, it is assumed that the temporal-spectral analysis (the last stage) is sufficient to characterize GCW calls and hence there is no uncertainty in this feature set.

%\begin{figure}[t!]
%	\centering
%	\begin{subfigure}[t]{0.5\textwidth}
%		\centering
%		\includegraphics[height=4cm]{\ROOT/phdthesis/cascadeReal.png}
%		\caption{The software components of the proposed sensor is organized as a cascade with 3 stages: per-frame analysis and detection as stage 1, multi-frame analysis and detection as stage 2, data logging and optimal signal detection as stage 3. Note that most of the cascade is implemented on the sensor, which is highly resource-constrained (see Section \ref{subsec:hardware}), except for the last stage decision, which is made on a client\footnotemark. Hence the system is distributed across the network, with the dashed line representing a remote connection.}
%		\label{fig:cascadeReal}
%	\end{subfigure}
%	~ 
%	\begin{subfigure}[t]{0.5\textwidth}
%		\centering
%		\includegraphics[height=2.8cm]{\ROOT/phdthesis/dutyReal.png}
%		\caption{The duty cycling design.}
%		\label{fig:dutyReal}
%	\end{subfigure}
%	\caption{}
%\end{figure}
%\footnotetext{The resource consumption on the client is intentionally left out since the focus is on the sensor only.}

\subsection{Results}

\begin{figure}[t!]
	\centering
	\includegraphics[width=\linewidth]{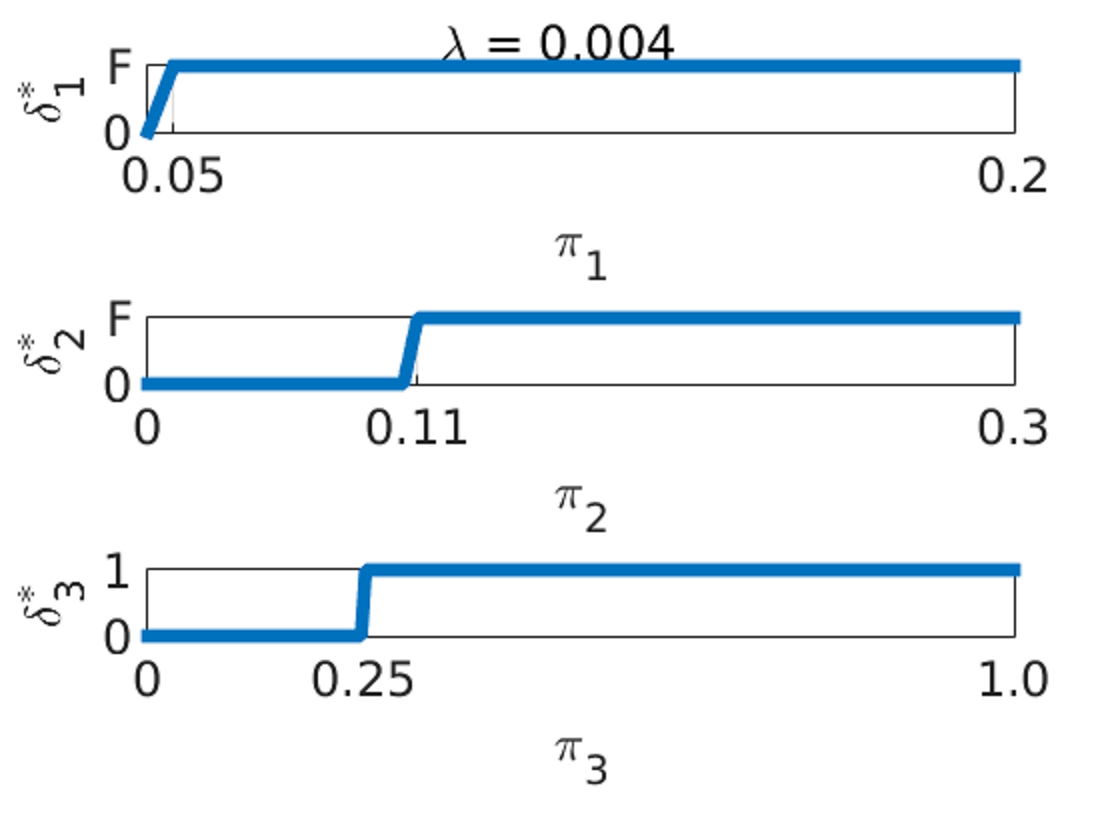}
	\caption{Optimal decision rules of the cascade system $\delta_i^{\ast}(\pi_i) \in \{F,0,1\}, i = 1,\dots,3$.}
	\label{fig:thresh}
\end{figure}

\begin{figure}[t!]
	\centering
	\includegraphics[width=\linewidth]{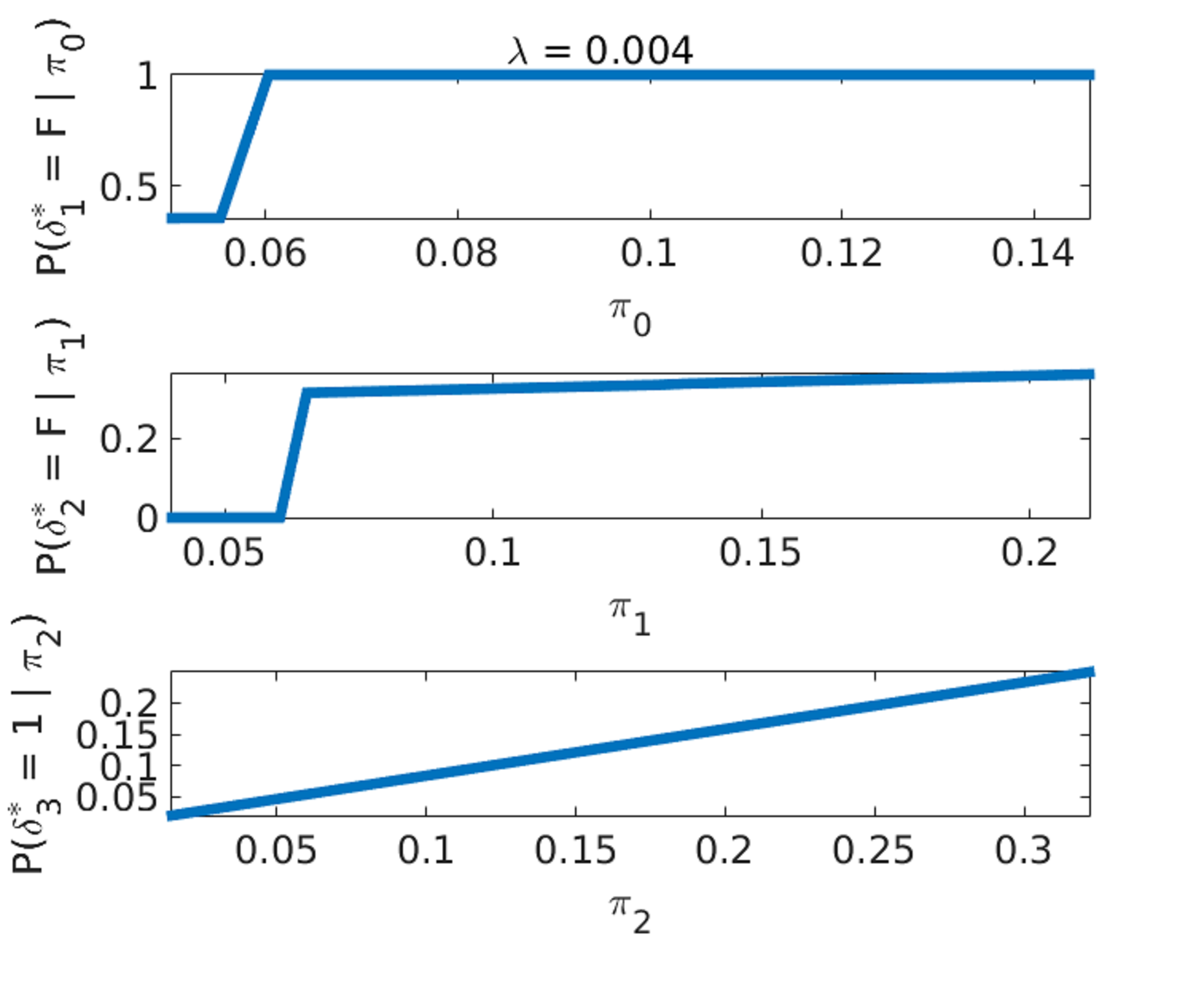}
	\caption{The alternative representation of the optimal solution for adaptive implementation.}
	\label{fig:act}
\end{figure}

The optimal thresholds/strategies for detectors in the cascade are given in Fig.\ \ref{fig:thresh}. The equivalent, implementation-friendly version of the solution, as discussed in Proposition \ref{prop:nonstatic}, is given in Fig.\ \ref{fig:act}. Note that the decision functions of intermediate layers have limited supports due to model uncertainties. Furthermore, Proposition \ref{prop:vivaCascade} can be applied to verify that there is no gain from having additional early positive decisions in this system.

%Overall, the optimized guided-processing system is $1.13\times$ (measured by the reduction in system risk, see Section \ref{subsec:compareDuty}) more energy-efficient than an equivalent system using duty-cycling at its best.
% Over the rare-target regime, both designs (the proposed and the best duty-cycling) achieve comparable miss ($1.47\%$ and $4.35\%$) and false-alarm (both $0.27\%$) rates. However, the proposed design achieves $2\times$ improvement in energy-saving/operational lifetime, i.e.\ an average energy consumption of $103.94$ mJ compared to $211.8$ mJ per audio frame of the best duty-cycling approach.

%(achieved by setting the regularization parameter $\lambda$ to $0.024$)
%Fig.\ \ref{fig:cascade-duty}, 

%\begin{figure}[t!]
%	\centering
%	\includegraphics[height=2.5in]{\ROOT/rmss/actProb}
%	\caption{The optimal thresholds at stage 1 and 2 control the probabilities of feature extraction, i.e.\ $\mathrm{P}(\delta_1(Y_1) = F|\pi_0) = \mathrm{P}(\pi_1(Y_1) \geq \tau_1^\ast|\pi_0)$ and $\mathrm{P}(\delta_2(Y_2) = F|\pi_1) = \mathrm{P}(\pi_2(Y_2) \geq \tau_2^\ast|\pi_1)$, respectively, and hence the system's resource consumption.}
%	\label{fig:actProb}
%\end{figure}

%\begin{figure}[t!]
%	\centering
%	\includegraphics[height=2.5in]{\ROOT/rmss/V3}
%	\caption{The value functions at all stages. If $\pi_0\in[0.1,0.2]$, then $\pi_{L1} = 0.017, \pi_{U1} = 0.260$, and $\pi_{L2} = 0.002, \pi_{U2} = 0.3654$.}
%	\label{fig:value}
%\end{figure}

\begin{figure}[t!]
	\centering
	\includegraphics[width=\linewidth]{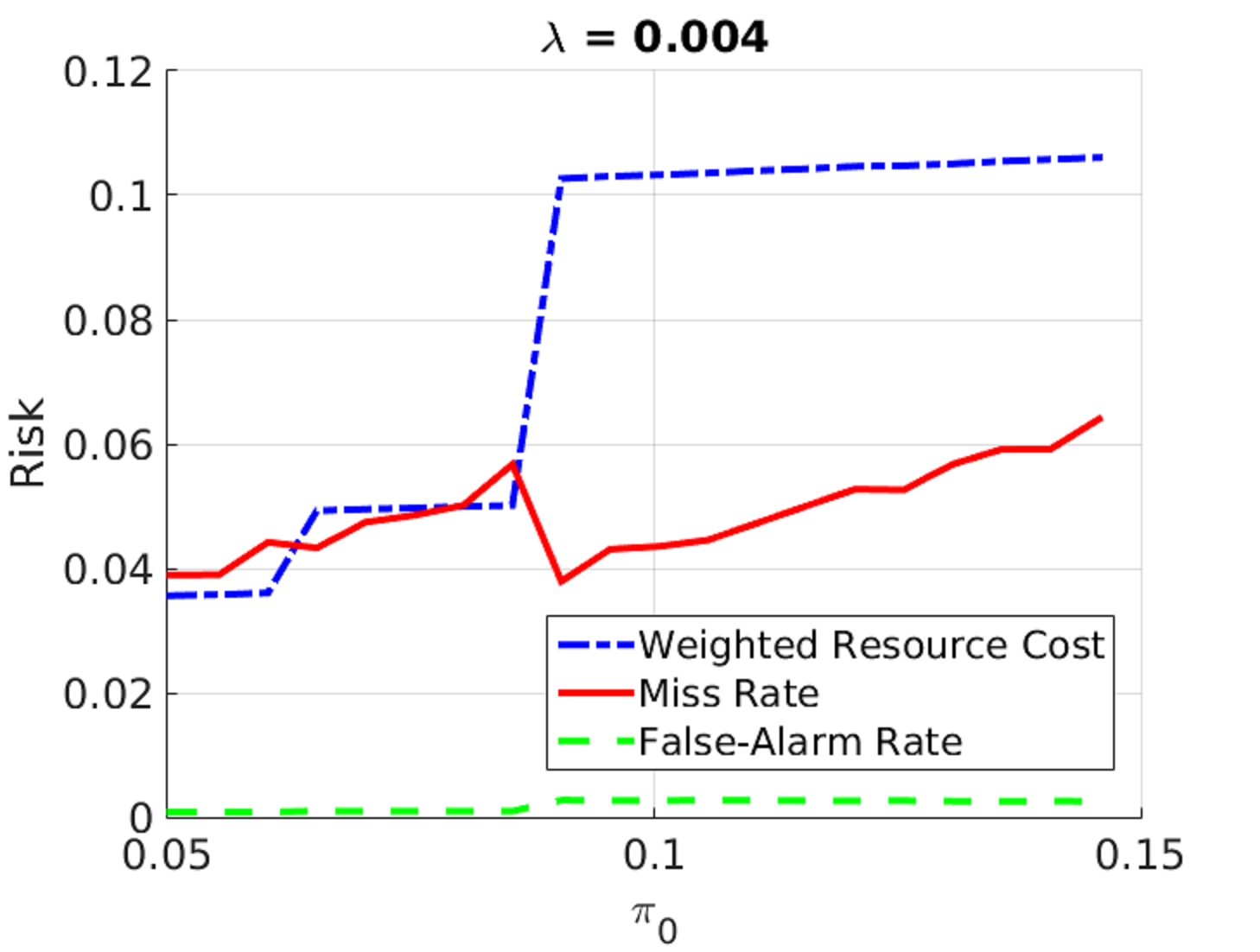}
	\caption{Breakdown of the system risk into components (see Eq.\ \eqref{eqn:minR}): false negative (miss), false positive (false-alarm), and Lagrangian-weighted resource consumption. Low false-alarm rate is achieved across the priors of interest. The miss rate tends to increase with the prior. At a certain level, the system must ramp up its resource consumption or incur more false-alarm to reduce the miss rate.}
	\label{fig:riskComponents3}
\end{figure}

The optimized system risk is further broken down into the weighted resource consumption, the miss and false-alarm rates in Fig.\ \ref{fig:riskComponents3} to provide an intuitive understanding of the optimal policies.

The guided-processing system is compared against both the theoretically-best (ideal) and the real, energy-equivalent duty-cycling designs, to be defined herein. In the ideal case, the lower bounds on energy costs and detection risks are assumed to hold, i.e.\
\begin{equation}
\begin{aligned}
&D_{\text{dc}} = D_K, d_{\text{dc}} = d_K\\
&R_{\text{dc},M} = R_{K,M}, R_{\text{dc},A} = R_{K,A}
\end{aligned}
\end{equation}
while the corresponding values in the real duty-cycling system must be measured directly. In addition, unlike the ideal case where it is sufficient to compare against $\rho \in \{1,0\}$ (either completely on or off, see Appendix \ref{subsec:app4}), $\rho$ must be adjusted in the real duty-cycling system to yield an equivalent energy consumption to the proposed one, thus allowing the two to be compared in term of their detection performance. 

Furthermore, to demonstrate the generalization power of the proposed approach over to that of \cite{jun2013cascading}, the system is also compared against its 2-stage version, where the spectral analysis in Fig.\ \ref{fig:cascadeReal} is removed (i.e.\ the client only executes the temporal-spectral analysis instead of a cascade of it and a spectral-analysis.).

\begin{figure}[t!]
	\centering
	\includegraphics[width=\linewidth]{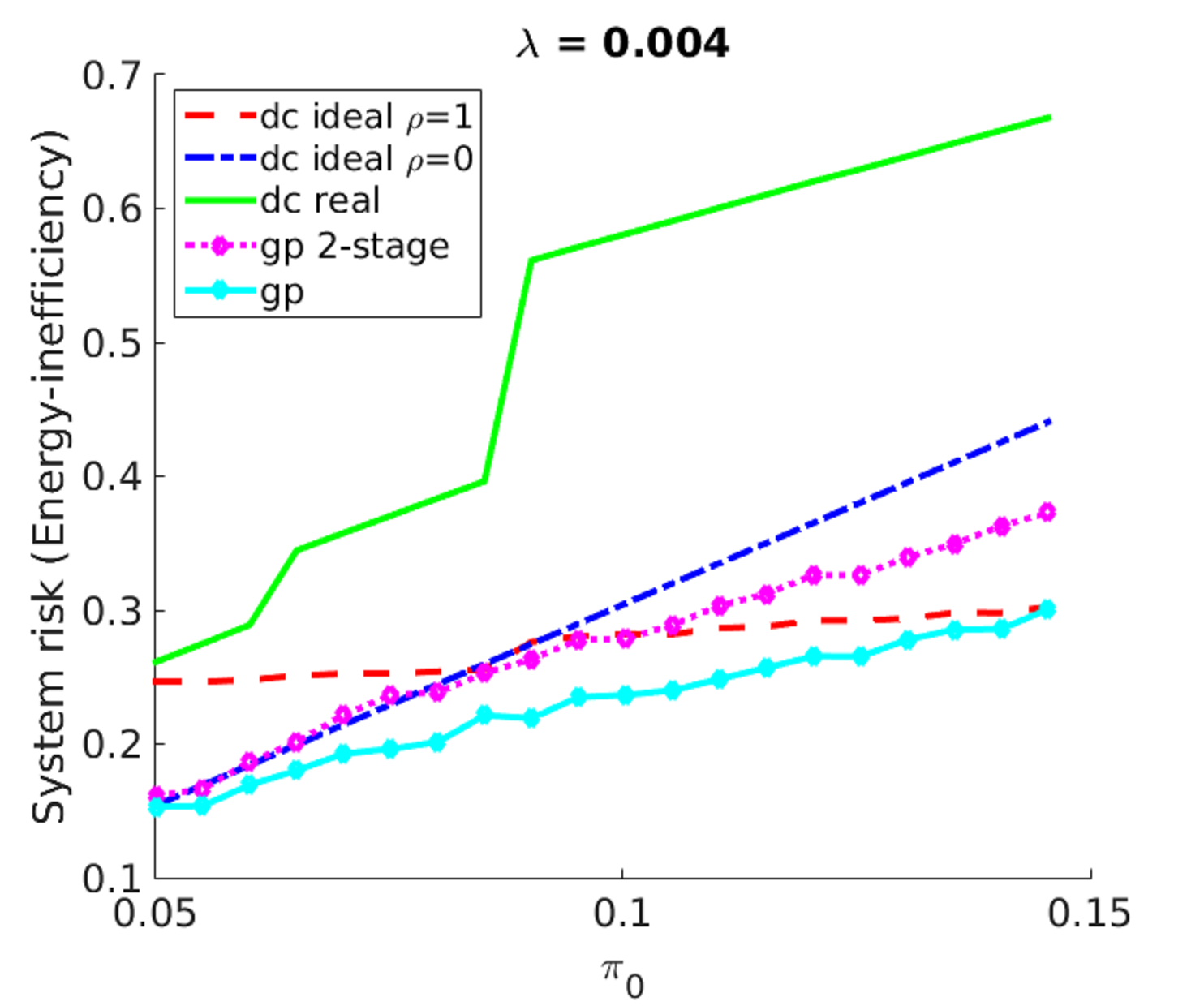}
	\caption{Comparison of system risk between the guided-processing (gp) and various duty-cycling (dc) approaches.}
	\label{fig:compRi}
\end{figure}

\begin{figure}[t!]
	\centering
	\includegraphics[width=\linewidth]{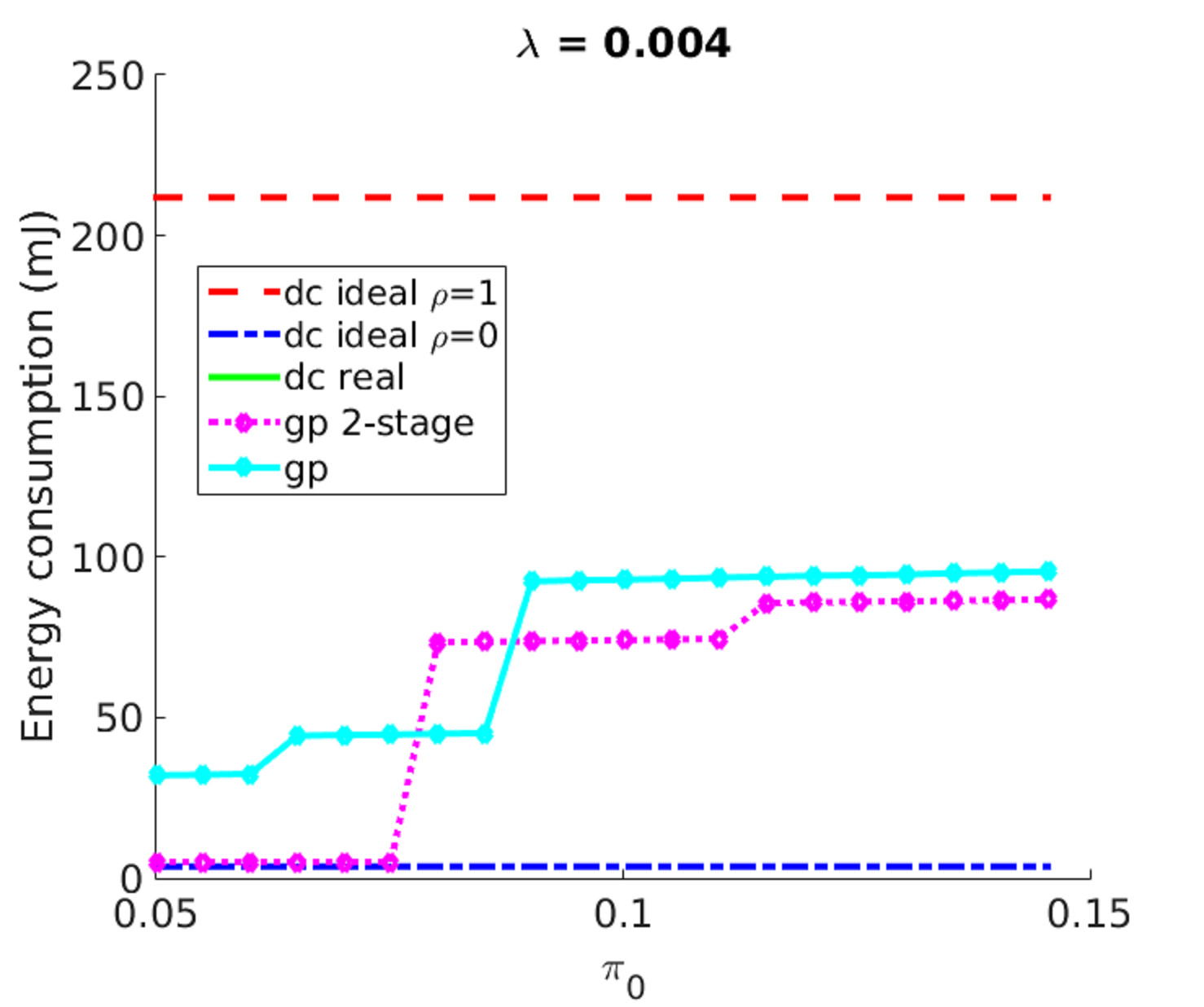}
	\caption{Comparison of energy consumption (per audio frame) between the guided-processing (gp) and various duty-cycling (dc) approaches. Note that the energy consumption of the real dc and gp approaches are the same by construction (i.e.\ their curves overlap by setting $\rho$ appropriately).}
	\label{fig:compEn}
\end{figure}

\begin{figure}[t!]
	\centering
	\includegraphics[width=\linewidth]{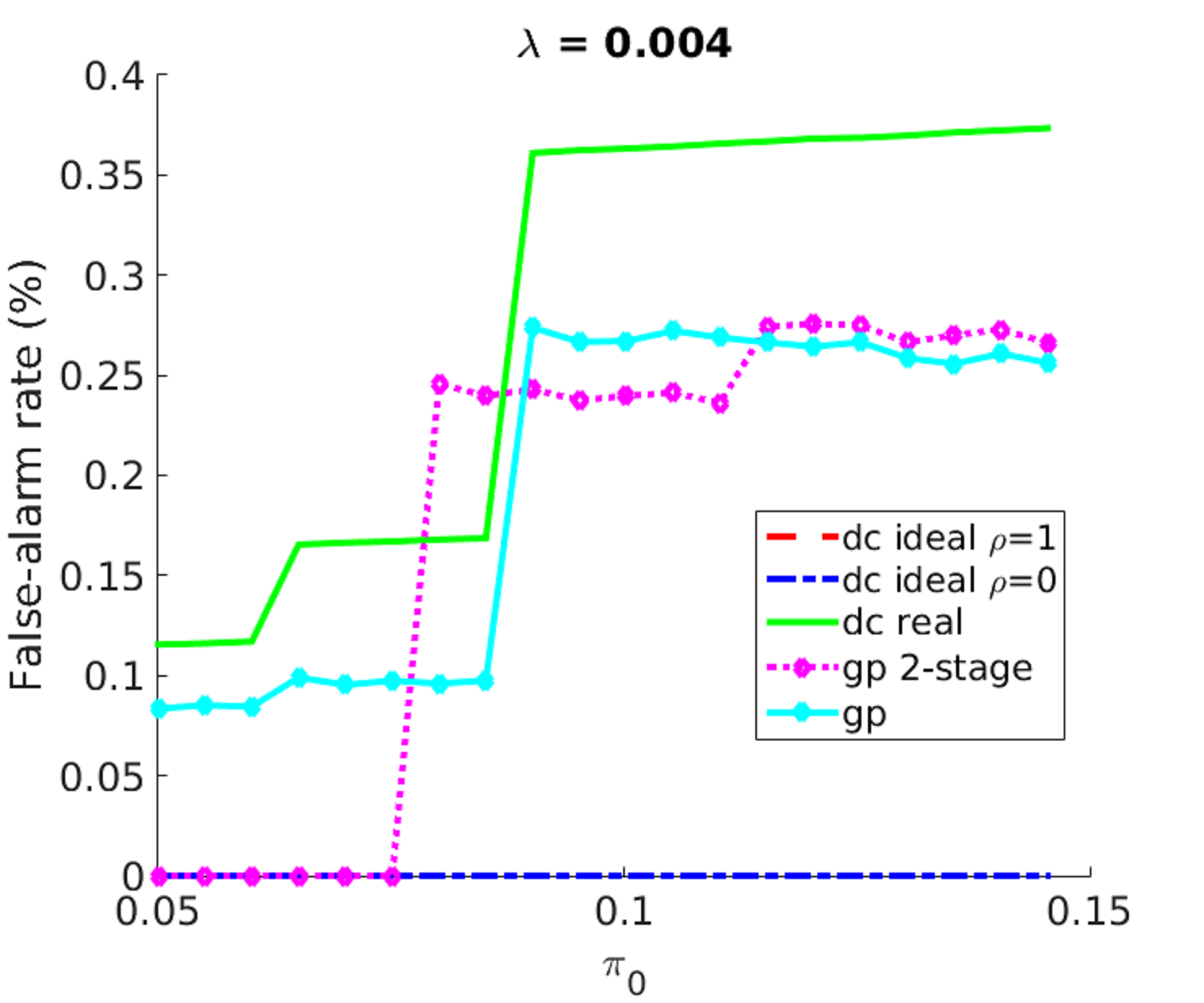}
	\caption{Comparison of false-alarm rate between the guided-processing (gp) and various duty-cycling (dc) approaches. Compared to dc real across $\pi_0$, gp is up to $1.7\times$ lower in false-alarm rate.}
	\label{fig:compFa}
\end{figure}

\begin{figure}[t!]
	\centering
	\includegraphics[width=\linewidth]{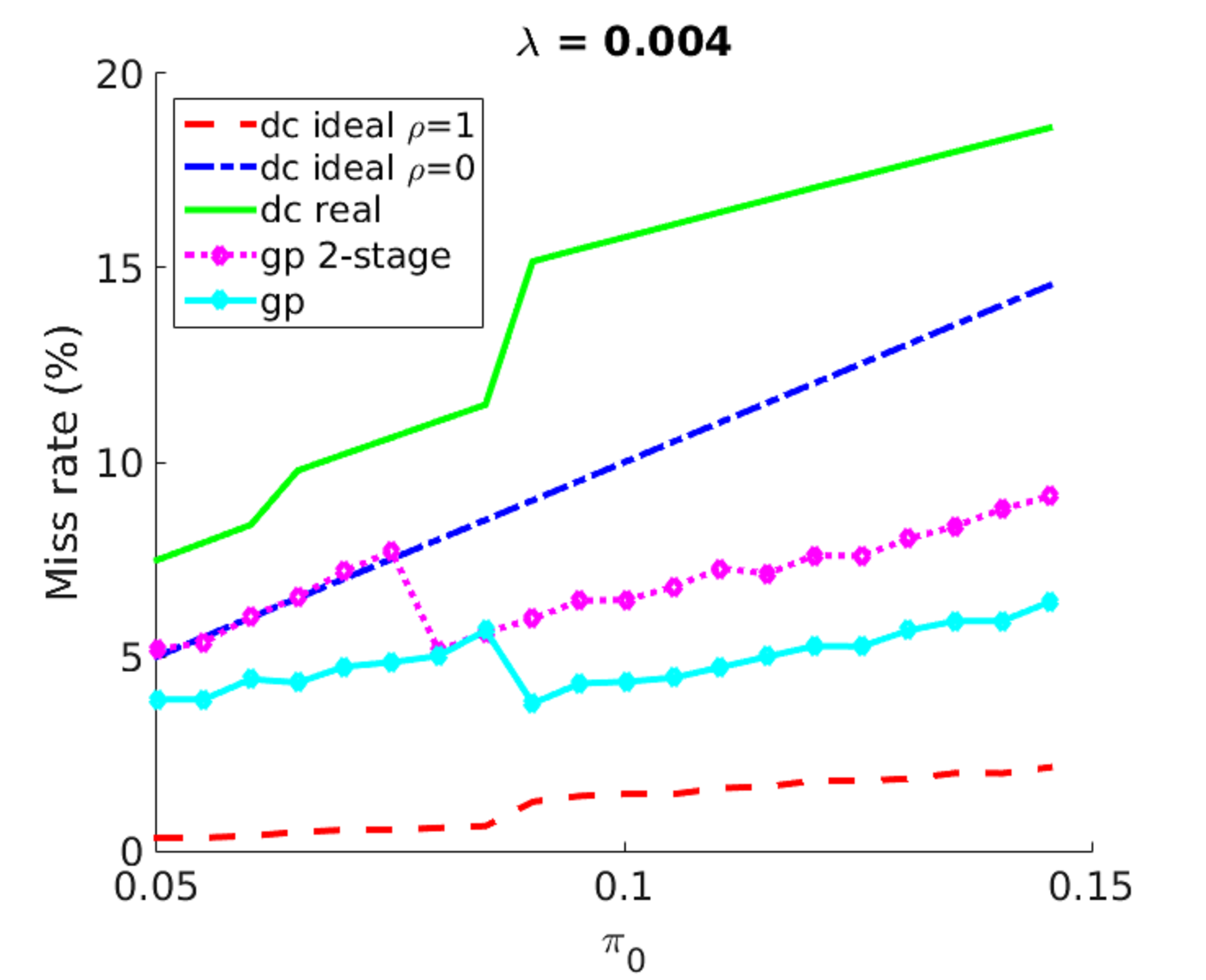}
	\caption{Comparison of miss rate between the guided-processing (gp) and various duty-cycling (dc) approaches. Compared to dc real across $\pi_0$, gp is up to $4\times$ lower in miss rate.}
	\label{fig:compMi}
\end{figure}

The comparison between the five approaches in term of system risk (energy-inefficiency), energy consumption, false-alarm and miss rates are given in Figures \ref{fig:compRi}, \ref{fig:compEn}, \ref{fig:compFa}, \ref{fig:compMi}, respectively. From Fig.\ \ref{fig:compRi}, it is evident that the proposed approach is the most energy-efficient one (with the smallest system risk) across the prior $\pi_0$ of interest. Moreover, the ideal bounds are tights, and generalization from two to three stages helps improves the overall energy-efficiency. Fig.\ \ref{fig:compEn}, \ref{fig:compFa}, \ref{fig:compMi} together show that the guided-processing approach is able to stay between the two ideal bounds for all three metrics and outperform the real duty-cycling approach in both false-alarm rate (up to $1.7\times$) and miss rate (up to $4\times$) for the same energy consumption.
Finally, it is worth noting that the removal of the spectral analysis module (resulting in the 2-stage version) strongly limits the design space and increases the total miss rate (even with one less miss term, see Fig.\ \ref{fig:compMi}) consistently across priors, when compared to the proposed 3-stage system.

\section{Conclusion}\label{sec:concl}

This paper proposes the guided-processing approach for sensing system design and shows that it can be fundamentally more energy-efficient than the naive approach of duty-cycling. Empirical evidence from a practical application also support the analysis. The proposed design was applied to develop an acoustic sensing service on which many applications can be built on top. These are publicly available online\footnote{At \url{http://acoustic.ifp.illinois.edu}} for demonstration.

An apparent drawback of the proposed approach is its stationary assumption and, as a result, the feedforward structure of the solution, i.e.\ the decision to invoke downstream processing rests entirely on an upstream detector with a fixed policy. It is conjecture that higher energy-efficiency can be achieved by exploiting the temporal structure in extracted features, for which a feedback-based solution might arise. For instance, it is natural for downstream results to influence upstream policies/decision-making over time.

%%%%%%%%%%%%%%%%%%%%%%%%%%%%%%%%%%%%%%%%%%%%%%%%%%%%
%%%%%%%%%%%%%%%%%%%%%%%%%%%%%%%%%%%%%%%%%%%%%%%%%%%%
\appendices

%\input{app}

\section*{Acknowledgements}
This work was supported in part by TerraSwarm, one of six centers of STARnet, a Semiconductor Research Corporation program sponsored by MARCO and DARPA,
and in part by a research grant for the Human Sixth
Sense Programme at the Advanced Digital Sciences Center from Singapore's Agency for Science, Technology and Research (A*STAR)


\appendix

\subsection{Proof of Theorem \ref{thm:optDecRules}}\label{subsec:app1}

We start by expanding the risk terms in \eqref{eqn:minR}. The false negative (miss) rate due to early negative decision for the first stage is 
\begin{equation}\label{eqn:multiR1}
\begin{aligned}
R_{1,M} &= \int \mathrm{p}(\mathrm{d}y_1)\big\{
C_M\pi_1(y_1)\mathbb{I}(\delta_1 = 0)
\big\}\\
\end{aligned}
\end{equation}
where $\mathbb{I}()$ denotes the indicator function that takes value $1$ when its argument is true and $0$ otherwise. $\mathrm{p}(\mathrm{d}y_{1:K})$ is the probability measure of feature realizations $y_{1:K}$.

Likewise, the miss terms for the stage $i=2,\dots,K$ can be given as follows.
\begin{equation}
\begin{aligned}
R_{i,M} &= \int \mathrm{p}(\mathrm{d}y_{1:i})\Big\{
C_M\pi_i(y_{1:i})\mathbb{I}(\delta_i = 0,\delta_{i-1} = F)\Big\}\\
%======
\end{aligned}
\end{equation}
Similarly, the false-alarm (false positive) term at the last stage is given as follows.
\begin{equation}
\begin{aligned}
R_{K,A} = &\int \mathrm{p}(\mathrm{d}y_{1:K})\Big\{
C_{A}(1-\pi_K(y_{1:K}))\\
&\mathbb{I}(\delta_K = 1,\delta_{K-1} = F)\Big\}\\
%======
\end{aligned}
\end{equation}

An important step in solving Problem \eqref{eqn:minR} is the following expansion of the expected resource cost in \eqref{eqn:lambdaE}. By the law of total probability,
\begin{equation}\label{eqn:totalProb}
\begin{aligned}
D_1 = D_1 \Big\{ \mathrm{P}(\delta_1=0) + \sum_{i=2}^{K-1} \mathrm{P}(\delta_{i}=0,\delta_{i-1}=F)+\\
\mathrm{P}(\delta_K=0,\delta_{K-1}=F)+\mathrm{P}(\delta_K=1,\delta_{K-1}=F) \Big\}\\
\end{aligned}
\end{equation}
and
\begin{equation}
\begin{aligned}
D_{i+1} \mathrm{P}(\delta_{i}=F) = D_{i+1} \Big\{ \sum_{j=i+1}^{K-1} \mathrm{P}(\delta_j=0,\delta_{j-1}=F)+\\
\mathrm{P}(\delta_K=0,\delta_{K-1}=F)+\mathrm{P}(\delta_K=1,\delta_{K-1}=F) \Big\},\\
i = 1,\dots,K-1
\end{aligned}
\end{equation}
Similar expansions can be done for $d_i,i=2,\dots,K$.

Putting everything back into \eqref{eqn:minR} yields a dynamic programming structure, with the state variable being the posteriors $\pi_i$ defined in Section \ref{subsec:featModel}. Minimizing \eqref{eqn:minR} can thus be achieved efficiently using the following backward procedure. 
\begin{equation}\label{eqn:VMultiApp}
\begin{aligned}
V_K(\pi_K) &\triangleq \min_{\delta_K}\mathbb{I}(\delta_K = 0)C_M\pi_K + \mathbb{I}(\delta_K = 1)C_{A}(1-\pi_K)\\
%=======
V_i(\pi_i) &\triangleq \min_{\delta_i}\mathbb{I}(\delta_i = 0) \left[ C_M\pi_i +\lambda{\bf d}_{i+1} \right]+\\
& \mathbb{I}(\delta_i = F) \left\{ \lambda D_{i+1}+\mathbb{E}[V_{i+1}(\pi_{i+1}(Y_{i+1},\pi_i))] \right\}\\
&i = 1,\dots,K-1\\
%=======
V_0(\pi_0) &\triangleq \lambda D_{1}+\mathbb{E}[V_{1}(\pi_1(Y_{1},\pi_0))]\\
\end{aligned}
\end{equation}
where the expectation is taken with respect to the evidence probabilities (see Section \ref{subsec:featModel})
%\begin{equation}
%\mathrm{p}_{i+1}(y_{i+1}|X_{i+1}=1)\pi_i + %\mathrm{p}_{i+1}(y_{i+1}|X_{i+1}=0)(1-\pi_i)
%\end{equation}
and $V_i$ is the value function at stage $i$. From the first and second expressions of \eqref{eqn:VMultiApp}, the minimizers for the system can be obtained by setting
\begin{equation}\label{eqn:deltaOpt1}
\delta_K^{\ast}(\pi_K) = 
\begin{cases}
0, \pi_K <C_A/(C_A+C_M)\\
1, \text{ else}
\end{cases}
\end{equation}
and
\begin{equation}\label{eqn:deltaOpt2}
\begin{aligned}
\delta_i^{\ast}(\pi_i) &= 
\begin{cases}
0, V_i(\pi_i) = C_M\pi_i + \lambda{\bf d}_{i+1}\\
F, V_i(\pi_i) < C_M\pi_i + \lambda{\bf d}_{i+1}
\end{cases},\\
&i = 1,\dots,K-1
\end{aligned}
\end{equation}
The expression in \eqref{eqn:deltaOpt2} can be further simplified into
\eqref{eqn:optDecRules} using Lemmas \ref{lem:EVconcave} and \ref{lem:EV0}.

\begin{lemma}\label{lem:EVconcave}
	$\mathbb{E}[V_{i+1}(\pi_{i+1}(Y_{i+1}, \pi))],i = 0,\dots,K-1$ and $V_i(\pi),i=1,\dots,K$ are concave\footnote{Moreover, $V_i(\pi),i=1,\dots,K$ can be shown to be piece-wise linear and concave, which was first observed and proven (by induction) in \cite[Smallwood and Sondik]{smallwood1973optimal}.}.
	\begin{proof}
		$V_K(\pi)$ is concave. Hence, by Lemma \ref{lem:concave}, $\mathbb{E}[V_{K}(\pi_{K}(Y_{K}, \pi))]$ is concave.
		
		Assume that $V_{i+1}(\pi)$ is concave, thus $\mathbb{E}[V_{i+1}(\pi_{i+1}(Y_{i+1}, \pi))]$ is concave by Lemma \ref{lem:concave}, then
		\begin{equation}
		V_i(\pi) = \min \{(\pi),\lambda D_{i+1}+\mathbb{E}[V_{i+1}(\pi_{i+1}(Y_{i+1}, \pi))]
		\end{equation} 
		is also concave. Again, by Lemma \ref{lem:concave}, $\mathbb{E}[V_{i}(\pi_{i}(Y_{i}, \pi))]$ is concave.
	\end{proof}
\end{lemma}

\begin{lemma}\label{lem:concave}
	$\mathbb{E}[V_{i+1}(\pi_{i+1}(Y_{i+1}, \pi))]$ is concave if $V_{i+1}(\pi)$ is concave.
	\begin{proof}
		See \cite[p. 146]{bertsekas1976dynamic}.
	\end{proof}
\end{lemma}

\begin{lemma}\label{lem:EV0}
	$\mathbb{E}[V_{i+1}(\pi_{i+1}(Y_{i+1}, 0))] = \lambda {\bf d}_{i+2}, i = 0,\dots,K-2$ and $\mathbb{E}[V_{K}(\pi_{K}(Y_{K}, 0))] = 0$.
	\begin{proof}
		$V_{K}(0) = 0$, then $\mathbb{E}[V_{K}(\pi_{K}(Y_{K}, 0))] = V_{K}(0) = 0$ and
		\begin{equation}
			V_{K-1}(0) = \lambda \min \{d_K,D_{K}\} = \lambda d_K
		\end{equation}
		Hence, $\mathbb{E}[V_{K-1}(\pi_{K-1}(Y_{K-1}, 0))] = V_{K-1}(0)= \lambda d_K$
		
		Now assume that $\mathbb{E}[V_{i+1}(\pi_{i+1}(Y_{i+1}, 0))] = \lambda {\bf d}_{i+2}$, then
		\begin{equation}
		V_i(0) = \lambda \min \{{\bf d}_{i+1},D_{i+1}+{\bf d}_{i+2}\} = \lambda {\bf d}_{i+1}. 
		\end{equation}
		Hence, $\mathbb{E}[V_{i}(\pi_{i}(Y_{i}, 0))] = V_{i}(0) = \lambda {\bf d}_{i+1}$.
	\end{proof}
\end{lemma}

\subsection{Proof of Proposition \ref{prop:vivaCascade}}\label{subsec:app2}
Introducing (additional) early positive decisions to intermediate stages results in the following modification to the second expression of \eqref{eqn:VMultiApp}.
\begin{equation}
\begin{aligned}
V_i(\pi_i) &\triangleq \min_{\delta_i}\mathbb{I}(\delta_i = 0)\left[ C_M\pi_i +\lambda{\bf d}_{i+1} \right]+\\
&\mathbb{I}(\delta_i = 1) \left[ C_A(1-\pi_i)+\lambda{\bf d}_{i+1} \right] + \\
& \mathbb{I}(\delta_i = F) \Big\{\lambda D_{i+1}+\mathbb{E}[V_{i+1}(\pi_{i+1}(Y_{i+1},\pi_i))]\Big\}\\
&i = 1,\dots,K-1\\
\end{aligned}
\end{equation}
Therefore the positive decision is \textit{not} chosen by the optimal policy under the following circumstances.
\begin{equation}\label{eqn:51}
\begin{aligned}
\delta_i^{\ast} \neq 1 \text{ if } V_i < C_A(1-\pi_i) +\lambda{\bf d}_{i+1},\\
i=1,\dots,K-1
\end{aligned}
\end{equation}
Since $V_i$ is a concave function of $\pi_i$, \eqref{eqn:51} is equivalent to
\begin{equation}
\begin{aligned}
\delta_i^{\ast} \neq 1 \text{ if } \pi_i \leq \max\{\pi_i:V_i < C_A(1-\pi_i)+\lambda{\bf d}_{i+1}\}, \\
i=1,\dots,K-1
\end{aligned}
\end{equation}

Hence if \eqref{eqn:condVivaCascade} holds then the positive decisions are never chosen by the optimal policy, and therefore do not make any difference in the system performance.

\subsection{Proof of Proposition \ref{propo:1}}\label{subsec:app4}

Recall that the feature model used in the duty-cycling design is the same as that of the cascade's last stage, i.e.\ the best one. The corresponding miss risk is then given by
\begin{equation}
\begin{aligned}
R_{\text{dc},M} &\triangleq \int \mathrm{p}(\mathrm{d}y_{K})  C_{M} \pi_K(y_{K})\mathbb{I}(\delta_{\text{dc}} = 0)\\
& = \int \mathrm{p}(\mathrm{d}y_{1:K})  C_{M} \pi_K(y_{1:K}) \mathbb{I}(\delta_{\text{dc}} = 0)\\
&\geq R_{K,M}^\ast
\end{aligned}
\end{equation}
where $\delta_{\text{dc}}$ is the duty-cycling's detection strategy. The second line follows from the law of total probability and the third one holds by definition. Similarly for the false-alarm risk, i.e.\
\begin{equation}
R_{\text{dc},A} \geq R_{K,A}^\ast
\end{equation}
From \eqref{eqn:duty}, the duty-cycling system risk is lower-bounded by
\begin{equation}\label{eqn:dutyRisk}
\rho(R_{K,M}^\ast + R_{K,A}^\ast + \lambda D_{K}) + (1-\rho)(C_{M}\pi_0+ \lambda d_{K})
\end{equation}
assuming zero overhead for duty-cycling, i.e.\ $D_{\text{dc}} = D_{K}$ and $d_{\text{dc}} = d_{K}$.
Let $\Delta R$ denote the difference between \eqref{eqn:dutyRisk} and the cascade performance in \eqref{eqn:minR}. Notice that $\Delta R(\rho)$ is a linear function of $\rho$. Hence, for the cascade design to outperform the duty-cycling design uniformly ($\Delta R(\rho) \leq 0, \forall \rho$), then $\Delta R(0)\leq 0$ and $\Delta R(1)\leq 0$ must hold.

The inequality $\Delta R(0)\leq 0$ is equivalent to the following trivial condition on the cascade design
\begin{equation}\label{eqn:always-off}
R^\ast \leq C_M \pi_0+\lambda d_{K}
\end{equation}
which simply states that the minimal risk achievable by the cascade design must be lower than that of doing nothing (the right-hand side of \eqref{eqn:always-off}).

On the other hand, the inequality $\Delta R(1)\leq 0$ is equivalent to the following non-trivial condition on the cascade design
\begin{equation}\label{eqn:rawCond}
\lambda E^\ast + \sum_{i=1}^{K-1}R_{i,M}^\ast \leq \lambda D_{K}
\end{equation}
Note that the optimal resource consumption $E^\ast$ is equal to the resource budget $e$. Therefore, \eqref{eqn:rawCond} is equivalent to \eqref{eqn:condBetter2}. The computation of $\sum_{i=1}^{K-1} R_{i,M}^\ast$ follows directly from Appendix \ref{subsec:app1}.

%\begin{equation}\label{eqn:E_ast}
%E^\ast = \frac{1}{\lambda } V_{0,E}(\pi_0)
%\end{equation}
%where $V_{0,E}(\pi_0)$ is the result of the following recursion
%\begin{equation}\label{eqn:ViE}
%\begin{aligned}
%V_{K,E}(\pi_K) &\triangleq 0,\pi_K\in[0,1]\\
%V_{i,E}(\pi_{i}) &\triangleq \begin{cases}
%0, &\pi_i\leq \tau_i^\ast\\
%\lambda D_{i+1} + \mathbb{E}[V_{i+1,E}(Y_{i+1},\pi_{i})], &\text{ else }\\
%\end{cases},\\
%&\pi_i\in [\pi_{Li},\pi_{Ui}], i = 1,\dots, K-1\\
%V_{0,E}(\pi_{0}) &\triangleq \lambda D_{1} + \mathbb{E}[V_{1,E}(Y_{1},\pi_{0})] \\
%\end{aligned}
%\end{equation}



\bibliographystyle{IEEEtran}
\bibliography{main}

\begin{thebibliography}{10}
\providecommand{\url}[1]{#1}
\csname url@samestyle\endcsname
\providecommand{\newblock}{\relax}
\providecommand{\bibinfo}[2]{#2}
\providecommand{\BIBentrySTDinterwordspacing}{\spaceskip=0pt\relax}
\providecommand{\BIBentryALTinterwordstretchfactor}{4}
\providecommand{\BIBentryALTinterwordspacing}{\spaceskip=\fontdimen2\font plus
\BIBentryALTinterwordstretchfactor\fontdimen3\font minus
  \fontdimen4\font\relax}
\providecommand{\BIBforeignlanguage}[2]{{%
\expandafter\ifx\csname l@#1\endcsname\relax
\typeout{** WARNING: IEEEtran.bst: No hyphenation pattern has been}%
\typeout{** loaded for the language `#1'. Using the pattern for}%
\typeout{** the default language instead.}%
\else
\language=\csname l@#1\endcsname
\fi
#2}}
\providecommand{\BIBdecl}{\relax}
\BIBdecl

\bibitem{atzori2010internet}
L.~Atzori, A.~Iera, and G.~Morabito, ``{The Internet of Things: A survey},''
  \emph{Computer networks}, vol.~54, no.~15, pp. 2787--2805, 2010.

\bibitem{mcafee2012big}
A.~McAfee, E.~Brynjolfsson, T.~H. Davenport, D.~Patil, and D.~Barton, ``Big
  data,'' \emph{The management revolution. Harvard Bus Rev}, vol.~90, no.~10,
  pp. 61--67, 2012.

\bibitem{viola2001rapid}
P.~Viola and M.~Jones, ``Rapid object detection using a boosted cascade of
  simple features,'' in \emph{Proceedings of the 2001 IEEE Computer Society
  Conference on Computer Vision and Pattern Recognition, 2001}, vol.~1.\hskip
  1em plus 0.5em minus 0.4em\relax IEEE, 2001, pp. I--511.

\bibitem{turaga2006resource}
D.~S. Turaga, O.~Verscheure, U.~V. Chaudhari, and L.~D. Amini, ``Resource
  management for networked classifiers in distributed stream mining systems,''
  in \emph{Sixth International Conference on Data Mining, 2006}.\hskip 1em plus
  0.5em minus 0.4em\relax IEEE, 2006, pp. 1102--1107.

\bibitem{le2013energy}
L.~Le, D.~M. Jun, and D.~L. Jones, ``Energy-efficient detection system in
  time-varying signal and noise power,'' in \emph{IEEE International Conference
  on Acoustics, Speech and Signal Processing (ICASSP), 2013}.\hskip 1em plus
  0.5em minus 0.4em\relax IEEE, 2013, pp. 2736--2740.

\bibitem{tang1991optimization}
Z.-B. Tang, K.~R. Pattipati, and D.~L. Kleinman, ``Optimization of detection
  networks: {Part I - Tandem structures},'' \emph{IEEE Transactions on Systems,
  Man and Cybernetics}, vol.~21, no.~5, pp. 1044--1059, 1991.

\bibitem{swaszek1993performance}
P.~F. Swaszek, ``On the performance of serial networks in distributed
  detection,'' \emph{IEEE Transactions on Aerospace and Electronic Systems},
  vol.~29, no.~1, pp. 254--260, 1993.

\bibitem{viswanathan1988optimal}
R.~Viswanathan, S.~C. Thomopoulos, and R.~Tumuluri, ``Optimal serial
  distributed decision fusion,'' \emph{IEEE Transactions on Aerospace and
  Electronic Systems}, vol.~24, no.~4, pp. 366--376, 1988.

\bibitem{luo2005optimization}
H.~Luo, ``Optimization design of cascaded classifiers,'' in \emph{IEEE Computer
  Society Conference on Computer Vision and Pattern Recognition, 2005},
  vol.~1.\hskip 1em plus 0.5em minus 0.4em\relax IEEE, 2005, pp. 480--485.

\bibitem{jun2010energy}
D.~M. Jun and D.~L. Jones, ``An energy-aware framework for cascaded detection
  algorithms,'' in \emph{2010 IEEE Workshop on Signal Processing Systems
  (SIPS)}.\hskip 1em plus 0.5em minus 0.4em\relax IEEE, 2010, pp. 1--6.

\bibitem{jun2013cascading}
------, ``{Cascading Signal-Model Complexity for Energy-Aware Detection},''
  \emph{IEEE Journal on Emerging and Selected Topics in Circuits and Systems},
  vol.~3, no.~1, pp. 65--74, 2013.

\bibitem{chen2016dual}
Y.~Chen, M.~Cho, S.~Jeong, D.~Blaauw, D.~Sylvester, and H.~S. Kim, ``A
  dual-stage, ultra-low power acoustic event detection system,'' \emph{IEEE
  International Workshop on Signal Processing Systems (SiPS)}, 2016.

\bibitem{chen2012fidelity}
J.~Chen, R.~Tan, G.~Xing, X.~Wang, and X.~Fu, ``Fidelity-aware utilization
  control for cyber-physical surveillance systems,'' \emph{IEEE Transactions on
  Parallel and Distributed Systems}, vol.~23, no.~9, pp. 1739--1751, 2012.

\bibitem{cohen2013managing}
D.~Cohen, ``Managing resources on a multi-modal sensing device for energy-aware
  state estimation,'' Master's thesis, 2013.

\bibitem{raykar2010designing}
V.~C. Raykar, B.~Krishnapuram, and S.~Yu, ``Designing efficient cascaded
  classifiers: tradeoff between accuracy and cost,'' in \emph{Proceedings of
  the 16th ACM SIGKDD International Conference on Knowledge Discovery and Data
  Mining}.\hskip 1em plus 0.5em minus 0.4em\relax ACM, 2010, pp. 853--860.

\bibitem{chen2012classifier}
M.~Chen, K.~Q. Weinberger, O.~Chapelle, D.~Kedem, and Z.~Xu, ``Classifier
  cascade for minimizing feature evaluation cost,'' in \emph{International
  Conference on Artificial Intelligence and Statistics}, 2012, pp. 218--226.

\bibitem{emre1999polarimetric}
B.~Emre~Ertin, ``Polarimetric processing and sequential detection for automatic
  target recognition systems,'' Ph.D. dissertation, The Ohio State University,
  1999.

\bibitem{trapeznikov2013multi}
K.~Trapeznikov, V.~Saligrama, and D.~Casta{\~n}{\'o}n, ``Multi-stage classifier
  design,'' \emph{Machine learning}, vol.~92, no. 2-3, pp. 479--502, 2013.

\bibitem{trapeznikov2013supervised}
K.~Trapeznikov and V.~Saligrama, ``Supervised sequential classification under
  budget constraints,'' in \emph{Proceedings of the Sixteenth International
  Conference on Artificial Intelligence and Statistics}, 2013, pp. 581--589.

\bibitem{wang2014lp}
J.~Wang, K.~Trapeznikov, and V.~Saligrama, ``An {LP} for sequential learning
  under budgets.'' in \emph{AISTATS}, 2014, pp. 987--995.

\bibitem{jun2013cheap}
D.~M. Jun, L.~Le, and D.~L. Jones, ``Cheap noisy sensors can improve activity
  monitoring under stringent energy constraints,'' in \emph{Global Conference
  on Signal and Information Processing (GlobalSIP), 2013 IEEE}.\hskip 1em plus
  0.5em minus 0.4em\relax IEEE, 2013, pp. 683--686.

\bibitem{huber1968robust}
P.~J. Huber, ``Robust confidence limits,'' \emph{Zeitschrift f{\"u}r
  Wahrscheinlichkeitstheorie und verwandte Gebiete}, vol.~10, no.~4, pp.
  269--278, 1968.

\bibitem{huber2011robust}
------, ``Robust statistics,'' \emph{International Encyclopedia of Statistical
  Science}, pp. 1248--1251, 2011.

\bibitem{levy2008principles}
B.~C. Levy, \emph{Principles of signal detection and parameter
  estimation}.\hskip 1em plus 0.5em minus 0.4em\relax Springer, 2008.

\bibitem{poor1998quickest}
H.~V. Poor \emph{et~al.}, ``Quickest detection with exponential penalty for
  delay,'' \emph{The Annals of Statistics}, vol.~26, no.~6, pp. 2179--2205,
  1998.

\bibitem{mor2016toward}
N.~Mor, B.~Zhang, J.~Kolb, D.~S. Chan, N.~Goyal, N.~Sun, K.~Lutz, E.~Allman,
  J.~Wawrzynek, E.~A. Lee \emph{et~al.}, ``Toward a global data
  infrastructure,'' \emph{IEEE Internet Computing}, vol.~20, no.~3, pp. 54--62,
  2016.

\bibitem{chodorow2013mongodb}
K.~Chodorow, \emph{MongoDB: the definitive guide}.\hskip 1em plus 0.5em minus
  0.4em\relax " O'Reilly Media, Inc.", 2013.

\bibitem{trepn}
L.~Ben-Zur, ``{Developer tool Spotlight - Using Trepn Profiler for
  Power-Efficient Apps},''
  \url{https://developer.qualcomm.com/blog/developer-tool-spotlight-using-trepn-profiler-power-efficient-apps},
  2011, [Online; accessed Oct-2014].

\bibitem{leonard2010variation}
W.~J. Leonard, J.~Neal, and R.~Ratnam, ``Variation of {Type B} song in the
  endangered {Golden-cheeked Warbler} ({Dendroica} chrysoparia),'' \emph{The
  Wilson Journal of Ornithology}, vol. 122, no.~4, pp. 777--780, 2010.

\bibitem{smallwood1973optimal}
R.~D. Smallwood and E.~J. Sondik, ``The optimal control of partially observable
  {Markov} processes over a finite horizon,'' \emph{Operations Research},
  vol.~21, no.~5, pp. 1071--1088, 1973.

\bibitem{bertsekas1976dynamic}
D.~P. Bertsekas, ``Dynamic programming and stochastic control,'' 1976.

\end{thebibliography}
%\%begin{thebibliography}{1}
%\end{thebibliography}


\balance
\begin{IEEEbiography}[{\includegraphics[width=1in,height=1.25in,clip,keepaspectratio]{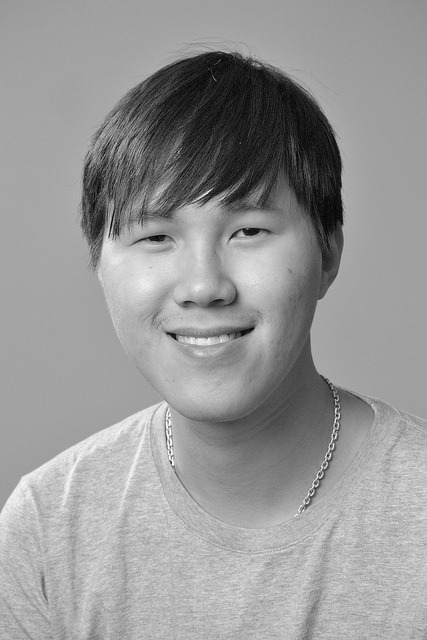}}]{Long N. Le}
	received the BSEE and MSEE degrees in
	electrical engineering in 2011 and 2013 from the Bach Khoa University and the University of Illinois at Urbana-Champaign, respectively.
	During the summer of 2015, he was an intern at the Audio and Acoustics Research Group at Microsoft Research.
	He is currently working toward the Ph.D. degree as a research assistant at the
	Coordinated Science Laboratory and Beckman Institute of the University of Illinois at Urbana-Champaign. 
	His current research interests include resource-efficient statistical inference and signal processing, with a focus on IoT applications.
\end{IEEEbiography}
%\balance
\begin{IEEEbiography}[{\includegraphics[width=1in,height=1.25in,clip,keepaspectratio]{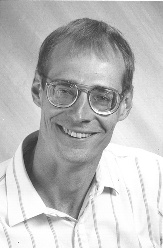}}]{Douglas L. Jones}
	received the BSEE, MSEE,
	and Ph.D. degrees from Rice University, Houston,
	TX, USA, in 1983, 1986, and 1987, respectively.
	During the 1987-1988 academic year, he was at
	the University of Erlangen-Nuremberg, Germany,
	on a Fulbright postdoctoral fellowship. Since 1988,
	he has been with the University of Illinois, Urbana-Champaign, IL, USA, where he is currently the director of Advanced Digital Sciences Center (ADSC) and a 
	Professor in Electrical and Computer Engineering,
	Neuroscience, the Coordinated Science Laboratory,
	and the Beckman Institute. He was on sabbatical
	leave at the University of Washington in Spring 1995 and at the University of
	California at Berkeley in Spring 2002. In the Spring semester of 1999 he served
	as the Texas Instruments Visiting Professor at Rice University. His research
	interests are in digital signal processing and systems, including nonstationary
	signal analysis, adaptive processing, multisensor data processing, OFDM,
	and various applications such as low-power implementations, biology and
	neuroengineering, and advanced hearing aids and other audio systems. He is
	an author of two DSP laboratory textbooks.
	Dr. Jones served on the Board of Governors of the IEEE Signal Processing
	Society from 2002 to 2004. He was selected as the 2003 Connexions Author of
	the Year.

\end{IEEEbiography}


%\newpage


%\vfill


\end{document}